\documentclass[3p,twocolumn]{elsarticle}

\usepackage[first,bottomafter,timestamp]{draftcopy}
\usepackage{mathrsfs}
\usepackage{hyperref,graphicx,amsfonts,amssymb,amsthm,amsmath,psfrag}
\usepackage{bm}       
\usepackage{color}
\usepackage{amssymb}
\usepackage{amsmath}     
\usepackage{hyperref} 
\usepackage{psfrag}
\usepackage[cp1250]{inputenc}
\usepackage{graphicx}
\usepackage{nccmath}
\usepackage[export]{adjustbox}
\DeclareMathAlphabet{\mathpzc}{OT1}{pzc}{m}{it}
\usepackage{booktabs}

\usepackage{comment}

\newcommand{\be}{\begin{equation}}
\newcommand{\ee}{\end{equation}}
\newcommand{\ba}{\begin{eqnarray}}
\newcommand{\ea}{\end{eqnarray}}

\begin{document}

\title{\boldmath Emergent scale symmetry: Connecting inflation and dark energy}
\author[heidelberg]{Javier Rubio} 
\ead{j.rubio@thphys.uni-heidelberg.de}
\author[heidelberg]{Christof Wetterich} 
\ead{c.wetterich@thphys.uni-heidelberg.de}

\address[heidelberg]{Institut f\"ur Theoretische Physik, Ruprecht-Karls-Universit\"at Heidelberg\\
Philosophenweg 16, 69120 Heidelberg, Germany.}

\begin{abstract}
Quantum gravity computations suggest the existence of an ultraviolet and an infrared fixed point where quantum 
scale invariance emerges as an exact symmetry. We discuss a particular \textit{variable gravity} model for the crossover between 
these fixed points which can naturally account for inflation and dark energy, using a single scalar field. In the Einstein-frame formulation the potential can be 
expressed in terms  of Lambert functions, interpolating between a power-law inflationary potential and a 
mixed-quintessence potential. For two natural heating scenarios, the transition between inflation and radiation
domination proceeds through a ``graceful reheating'' stage.  The radiation temperature significantly 
exceeds the temperature of big bang nucleosynthesis. For this type of model, the observable consequences
of the heating process can be summarized in a single parameter, the heating efficiency. Our quantitative analysis 
of compatibility with cosmological observations reveals the existence of realistic models able to describe the whole 
history of the Universe using only a single metric and scalar field and involving just a small number of order one 
parameters.
\end{abstract}

\maketitle

\section{Introduction}\label{Introduction}  

A dynamical scalar field with a sufficiently flat potential and at most tiny couplings to ordinary matter is often advocated as a promising alternative to the cosmological
constant \cite{Wetterich:1987fm,Ratra:1987rm}. This idea, usually named quintessence,  can partly be viewed as a late-time implementation 
of the successful inflationary paradigm.  Since inflation and dark energy share many essential properties, it is 
natural to seek for a unification of these two mechanisms into a common framework 
\cite{Peebles:1998qn,Spokoiny:1993kt,Brax:2005uf}. In this paper, we postulate that inflation and dark energy are intimately related to an underlying symmetry: scale invariance. 

When dealing with scale invariance one can take two different perspectives: i) assume that scale invariance 
remains an exact  symmetry even when quantum corrections are taken into account~\cite{Shaposhnikov:2008xi} 
or ii) assume that scale invariance is broken by quantum effects but will be approximately realized 
\textit{close to fixed points} \cite{Wetterich:1987fm}. Cosmological 
models based on the first line of reasoning and their associated phenomenology can be found in 
Refs.~\cite{Fujii:1974bq}-\cite{Kannike:2016wuy}.
Cosmological models of the second type resulting in a dilatation anomaly that vanishes asymptotically in 
the infinite future led to the first proposal of dynamical dark energy or quintessence~\cite{Wetterich:1987fm,Wetterich:1994bg}.

In this work we will adopt the second point of view. In particular, we will assume that scale invariance
is generically broken by the conformal anomaly but it reemerges as an exact quantum symmetry  in the early-
and late-time evolution of the Universe. The resurgence of the symmetry can be related to the presence of 
ultraviolet (UV) and infrared (IR) fixed points in the renormalization group flow. In the vicinity of these
points, any information about the mass scales in the theory is lost \cite{Wetterich:2014gaa}.  This idea 
can be easily implemented in a variable gravity scenario \cite{Wetterich:2014gaa,Wetterich:2013jsa,
Wetterich:2013aca}.  

In this paper we present the complete cosmological history for a particular \textit{crossover variable gravity} model with a 
singlet scalar field. In the \textit{scaling frame} the field is coupled nonminimally to gravity and to the Standard Model, supplemented by some unspecified dark matter candidate
and potentially by heavy particles as in grand unification. 
The model contains no tiny or huge dimensionless quantities put in by hand. The four parameters appearing in 
the effective action are all of order one. The first three describe the approach to the UV and IR fixed 
points \textit{in the scalar sector} and the position on the crossover trajectory. The last parameter describes the present growth rate of neutrino masses, which is associated to the coupling between the scalar field and neutrinos. For early cosmology, the net effect of the interactions 
between the scalar field and the Standard Model particles (and possible sectors beyond that) can be summarized in 
a \textit{heating efficiency} $\Theta$. These few parameters are sufficient for a quantitative account of the history of the 
Universe from inflation to the present accelerated expansion era. Our simple model seems so far compatible 
with cosmological observations. Neither tiny nor fine tuned parameters are 
introduced to explain the small value of the present dark energy density, which is rather a consequence of 
the long age of the Universe in Planck units.

The comparison of our model with cosmological observations is performed in the Einstein frame with a canonical kinetic
term for the scalar field. This allows us to 
find explicit analytic solutions and facilitates the comparison with other quintessential inflation models in the literature,
see for instance Refs.~\cite{Peebles:1998qn,Spokoiny:1993kt,Brax:2005uf} for well-known examples 
and Refs.~\cite{Hossain:2014xha,Agarwal:2017wxo,Ahmad:2017itq,Geng:2017mic} for recent discussions. We follow here the general approach in which the inflationary  epoch is followed by
a transition to a scaling or tracker solution of which the long duration is responsible for the tiny value of the present dark 
energy density. The end
of this scaling era is triggered by neutrinos with growing masses 
that become nonrelativistic in the recent cosmological history. This general scenario, originally proposed in
Refs.~\cite{Wetterich:2013aca,Wetterich:2013wza}, has been recently followed by 
several groups \cite{Hossain:2014xha,Agarwal:2017wxo,Ahmad:2017itq,Geng:2017mic}. 
The explicit Einstein-frame formulation presented in this paper allows us to replace arguments for 
approximate solutions by exact analytical results, which substantially extends the range of validity of the scenario
in parameter space. 

Beyond the explicit and convenient solutions in the Einstein frame, our investigation contains several 
new results. We propose for the \textit{heating} or entropy production preceding the radiation dominated epoch a  general 
mechanism that is neither gravitational particle production nor instant preheating. Only 
the latter two mechanisms have been previously discussed within models of quintessential inflation. The mechanism 
presented in this paper is based on the general framework for particle production in the presence of time-varying fields, 
but adapted to the situation in which the potential does not have a minimum. The absence of a minimum is 
required for the transition from inflation to a tracker solution and typical for a variable gravity framework 
containing a single crossover at early times. For this scenario all features relevant for observations can be 
summarized into a single parameter -- the heating efficiency $\Theta$. The duration of the kination epoch between 
the end of inflation and the onset of the radiation dominated epoch can be rather short, leading to a high 
heating temperature. We find that the (almost massless) cosmon excitations generated during the heating stage
do not significantly contribute to the effective number of neutrino species at big bang nucleosynthesis.

This paper is organized as follows. In Section~\ref{sec:scenario}, we present the effective action of 
the model in a \textit{scaling frame} where the Planck mass is given by a scalar field. We describe the properties
of the UV and IR fixed points
responsible for the early- and late-time acceleration of the Universe. In Section~\ref{sec:scenario2}, we  
reformulate the variable gravity scenario into the more common, although completely equivalent, Einstein 
frame. This formulation is used in the following sections to study the cosmological  implications of the
model. Section~\ref{sec:inflation} contains the details of inflation. We show that the UV fixed point gives
rise to a power-law  Einstein-frame potential and derive the associated inflationary observables.  The spectral
tilt and the tensor-to-scalar ratio are shown to be related and to depend  only on the UV fixed-point anomalous 
dimension. The initial stages of the postinflationary dynamics are discussed in Section \ref{sec:kinetial}. The
crossover to the IR fixed point translates into the appearance of a field region where the Einstein-frame potential becomes 
steep. This triggers the onset of a kinetic domination regime. The kinetic regime must be limited in time 
for the model to be cosmologically viable. In particular, part of the energy density of the inflaton field must be transmitted 
to the Standard Model particles, which must become the dominant energy component  before big bang nucleosynthesis (BBN).  In 
Section~\ref{sec:reheating}, we discuss two natural heating mechanisms and determine the associated radiation
temperature (``reheating'' temperature).  We argue that a total decay of the inflaton field is not possible, and is
neither necessary nor even preferable. The evolution after heating 
and the onset of the dark energy dominated era are discussed in Section \ref{sec:hbb}. Section~\ref{sec:conclusions} 
contains our conclusions. Appendix A summarizes several properties of Lambert functions that are useful for the derivation of 
the analytic solutions presented in this paper. Appendixes B and C contain details of our heating scenario and of the creation 
of cosmon excitations during this period.

\section{Variable gravity scenario}\label{sec:scenario}

The variable gravity scenario is usually formulated in a \textit{scaling frame} in which not only the Planck scale, 
but also the dimensionless couplings and masses of elementary particles are allowed to depend on the 
expectation value of a scalar field $\chi$. We consider here a simple real scalar which plays simultaneously 
the role of the inflaton, the cosmon, or the dilaton. The effective Lagrangian density for the graviscalar sector of the theory 
reads~\cite{Wetterich:2014gaa,Wetterich:2013jsa,Wetterich:2013aca}
\be\label{actionJ}
\frac{\cal L}{\sqrt{- \tilde g}}=
\frac{\chi^2}{2} \tilde R-\frac{B(\chi/\mu)-6}{2}(\tilde \partial\chi)^2-\mu^2\chi^2\,,
\ee
where the tilde denotes quantities in the \textit{scaling frame} and we have suppressed Lorentz indices. The 
implicit contractions in this paper should  be understood in terms of the metric associated with the frame under 
consideration. 

 The cosmon field $\chi$ in Eq.~\eqref{actionJ} defines the effective variable Planck
mass.  We will see that for the cosmological solutions of the field equations derived from the action 
\eqref{actionJ} it increases with time, 
with $\chi(t\rightarrow -\infty)\rightarrow 0$ and $\chi(t\rightarrow \infty)\rightarrow \infty$. The only fixed scale not proportional to the cosmon field is the scale $\mu$, which 
is associated to the scale or dilatation anomaly. The value of $\mu$ has no intrinsic meaning and can be used to set the mass scales. We will take
\begin{equation}\label{muvalue}
 \mu^{-1}=10^{10}\,{\rm yr}=1.2\times 10^{60} \, M_P^{-1}\,.
\end{equation}
For this choice the present value of the variable Planck mass in Eq.~\eqref{actionJ} amounts to
$M_P=\chi(t_0)=2.48\times 10^{18}$ GeV \cite{Wetterich:2013aca}. In other words, the increasing ratio $\chi/\mu$ has reached today 
a value $1.2\times 10^{60}$.

We have chosen to normalize the scalar field by its coupling to curvature in the scaling frame, i.e. by the first term in the right 
hand side of Eq.~\eqref{actionJ}. With this normalization, the scalar kinetic term has typically a nonstandard 
normalization, as reflected by the dimensionless function $B(\chi/\mu)$. In order to have a well-defined kinetic term
during the whole
cosmological evolution, we will require the function $B$ to be a positive function of $\chi/\mu$.  For $\mu=0$ and 
constant $B$ the associated action is scale invariant, while for $B=\mu=0$  the action is also 
conformally invariant and the cosmon field $\chi$ no longer propagates. 

For the matter and radiation sectors we take the Standard Model of particle physics with possible extensions 
including dark matter. We assume that at large $\chi$ the values of all the (renormalizable) dimensionless couplings 
in the Standard Model become independent of $\chi$, as required by scale symmetry. In practice, this implies that 
the Fermi scale and the confinement scale of strong interactions are proportional to $\chi$. The masses and binding 
energies of all elementary particles are then proportional to the dilaton expectation value, while cross 
sections scale as $\chi^{-2}$. In consequence, our setting is compatible with the equivalence principle tests and the severe bounds on the variation of fundamental constants \cite{Uzan:2010pm}.

A recent quantum gravity computation based on functional renormalization has indeed found for variable gravity a 
quadratic increase of the scalar potential for large $\chi$ \cite{Wetterich:2017ixo}. A strong enhancement of the effect of long-distance graviton 
fluctuations avoids a potential instability of the graviton propagator that would arise for a potential increasing 
faster than $\chi^2$. More generally, large classes 
of effective actions containing no more than two derivatives can be brought to the form \eqref{actionJ} by 
appropriate nonlinear field redefinitions \cite{Wetterich:2014gaa}. For example, this concerns potentials 
of the form $V=\alpha\mu^4+\mu^2\chi^2$.

A given model is specified by a choice of $B(\chi/\mu)$. For successful quintessential inflation one needs large $B$ during the inflationary epoch and small 
$B$ after the end of inflation. Large $B$ ensures slow-roll dynamics during inflation, while inflation ends once $B$ gets small. In this paper, we will concentrate on a particular scenario where $B$ satisfies the flow equation 
\be\label{flowB}
\mu\frac{\partial B}{\partial \mu}=\frac{\kappa\sigma B^2}{\sigma+\kappa B}\,.
\ee
This equation contains an infrared fixed point $B_*=0$, approached for $B\to 0$ with a quadratic term
\be\label{GaussFP}
\mu\partial_\mu B=\kappa B^2\,.
\ee
The ultraviolet fixed point for $B\to \infty$ 
\be\label{UVFP}
\mu\partial_\mu B =\sigma B\,,
\ee
is characterized by an anomalous dimension $\sigma$.

No quantum gravity computation for the flow of $B$ is available so far. Eq.~\eqref{flowB}  should be therefore
understood as an educated guess, or an assumption, on the exact quantum gravity dynamics. As suggested by the first investigations
in Ref.~\cite{Henz:2016aoh}, we assume the renormalization flow of quantum gravity to admit both a UV and an IR fixed point. The enhanced
conformal symmetry for $B=\mu=0$ implies that the $\beta$-function for $B$ vanishes for $B=0$. If the 
$\beta$-function in the IR limit is analytic in $B$ around $B=0$, i.e. $B=\sigma_{\rm IR} B+\kappa B^2$, the assumption of a 
vanishing infrared anomalous dimension $\sigma_{\rm IR}=0$ motivates the limit \eqref{GaussFP}. A simple way of achieving
large $B$ in the UV limit is an anomalous dimension of the scalar wave function renormalization, leading to the limit
\eqref{UVFP}. The precise interpolation between the UV and IR fixed points in Eq.~\eqref{flowB} is not important for 
the observable consequences of the model. 
A reason for the selection of the particular crossover in Eq.~\eqref{flowB}  is its simplicity. The scalar-gravity 
sector contains only three 
order one parameters: two constants $\sigma$ and $\kappa$ and an integration constant $c_t$ selecting 
a particular trajectory in the flow. The resulting tensor-to-scalar ratio 
of primordial perturbations turns out to be comparatively large, $r\simeq 0.05-0.1$ \cite{Wetterich:2014gaa} (see 
also Ref.~\cite{Hossain:2014xha}). Smaller values 
of $r$ can be obtained by modifying the behavior of $B$ at small $\chi$, for example by assuming 
a fixed point of the flow at some large but finite $B_*$ \cite{Wetterich:2013wza}, instead of the limit \eqref{UVFP}.

Since the main points of 
this paper will not be affected by the details of the function $B$, we will take advantage of 
the simplicity of Eq.~\eqref{flowB} for finding explicit solutions. Indeed, Eq.~\eqref{flowB} can be easily
integrated to obtain 
\be\label{Brun0}
\frac{\sigma}{\kappa B} +\ln \frac{\sigma}{\kappa B}=\ln 
\left[\frac{\sigma}{\kappa}\left(\frac\chi m\right)^\sigma\right]\,,
\ee
or equivalently (cf. Eq.~\eqref{Lamdef})
\be\label{Brun}
\frac{\sigma}{\kappa B(\chi)} ={\cal W}\left[\frac{\sigma}{\kappa} \left(\frac{\chi}{m}\right)^{\sigma} \right]\,,
\ee
with ${\cal W}$ the Lambert function \cite{refLambert} and 
\begin{equation}\label{mdef}
m\equiv\mu \exp(c_t)\,,
\end{equation}
a \textit{crossover} scale related to the integration constant $c_t$ via dimensional transmutation.

\section{Einstein-frame formulation}\label{sec:scenario2}

Most of the literature on inflation and on dynamical dark energy employs  a canonically normalized scalar field
in the Einstein frame. In order to permit an easy access and comparison of models for a wider community, the 
investigations and results of the present paper will be performed in this setting. The transformation of our 
variable gravity scenario to the Einstein frame will be done in two steps. The first realizes the Einstein 
frame with a fixed Planck mass $M_P$ and a noncanonically normalized scalar field. The second step proceeds to a 
canonical normalization of the scalar kinetic term.

Performing a conformal transformation $ g_{\mu\nu}=(\mu^2/M_P^2) \,V^{-1}(\chi(\varphi))\tilde g_{\mu\nu}$, with 
dimensionless scalar potential
\be \label{Wvarphi}
V(\chi(\varphi))=\left(\frac{\mu}{\chi}\right)^2=e^{-\alpha\varphi/M_P}\,,
\ee
and reduced Planck mass $M_{P} = 2.435\times 10^{18}$ GeV, we obtain
\ba\label{EFaction}
\frac{\cal L}{\sqrt{- g}} =
\frac{M_P^2}{2} R-\frac12k^2(\varphi) (\partial \varphi)^2- M_P^4\,V(\varphi)\,, 
\ea
with
 \ba
&& \hspace{-1cm}k^2(\varphi)=\frac{\alpha^2}{4}B(\varphi) \label{kphi}\,, \\
&& \hspace{-1cm}B(\varphi) =\frac{\sigma}{\kappa} {\cal W }^{-1}\left[\frac{\sigma}{\kappa}
\left(\frac{m}{\mu}\right)^{-\sigma}\exp{\left(\frac{\alpha\sigma\varphi}{2M_P}\right)} \right]\,. \label{BW} 
\ea
 The constant $\alpha$ in Eq.~\eqref{kphi} can be chosen to get the standard normalization ($k^2=1$)  
 in the present cosmological epoch \cite{Wetterich:2013wza} 
\be
\alpha^2=\frac{4}{B(\chi=M_P)}\approx 4\kappa\ln(M_P/m)\,.
\ee
One could also take $\alpha=1$. As we will see below, the 
constant $\alpha$ will completely disappear after canonically normalizing the scalar kinetic term.

Due to the positive definite choice of $B$ in Eq.~\eqref{actionJ}, the Einstein-frame 
Lagrangian \eqref{EFaction} is ghost free. \noindent In this basis, the cosmon potential $V(\varphi)$ decays 
exponentially to zero~\cite{Wetterich:1987fm,Wetterich:1994bg} and the dynamical information 
is encoded in the \textit{kinetial} $k^2(\varphi)$ \cite{Wetterich:2013jsa,Wetterich:2013wza}, see also 
Ref.~\cite{Dimopoulos:2017zvq}. 
\begin{figure}
\centering
\includegraphics[scale=0.37]{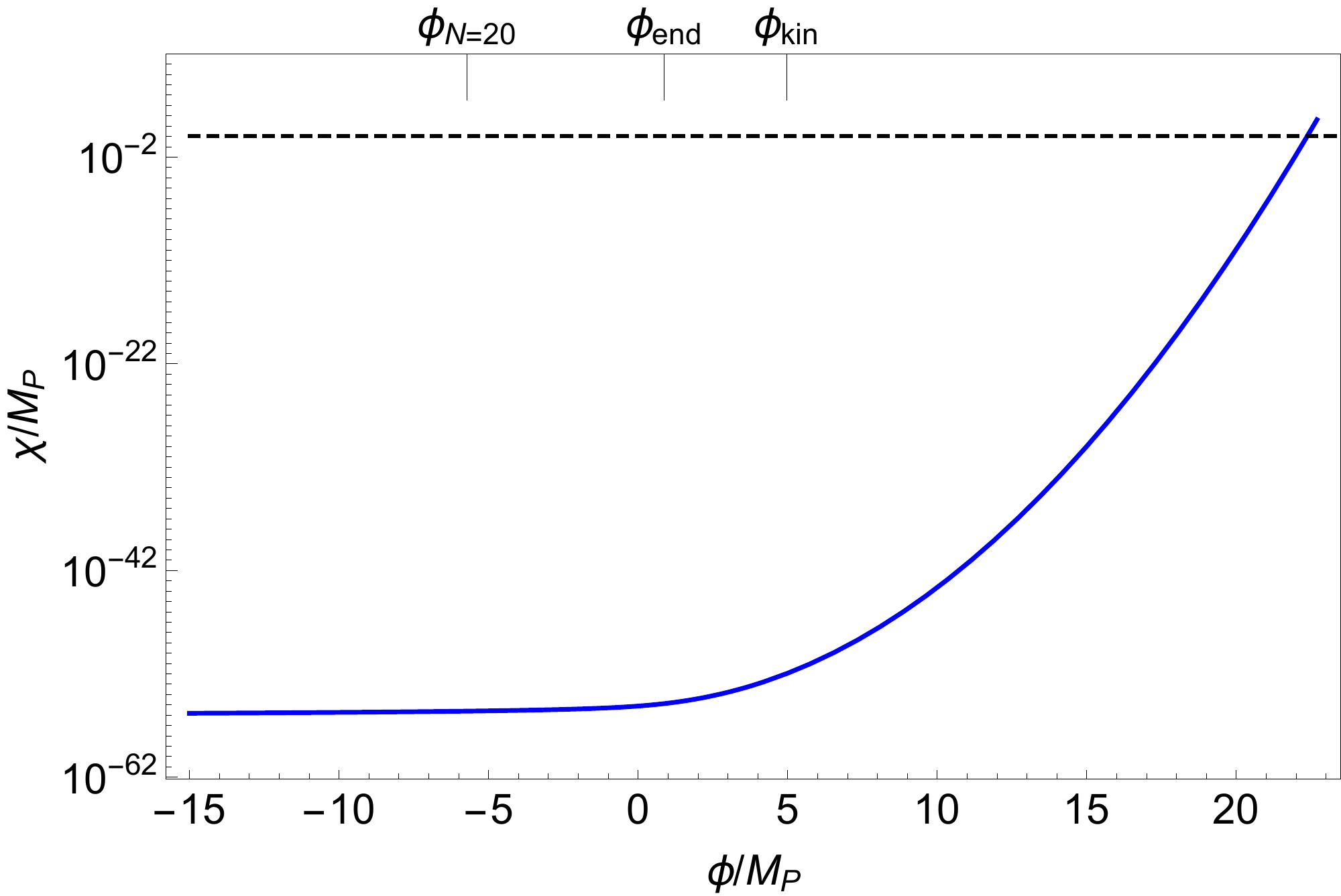}
\caption{The relation between $\chi$ and $\phi$ given by Eq.~\eqref{Wexact}. For this figure we took 
$\sigma=4$, $\kappa=1$ and $m=10^{5}\mu$ with $\mu$ given by Eq.~\eqref{muvalue}. The black-dashed 
line corresponds to $\chi=M_P$. For reference, we indicate the values of the field $\phi$ (in $M_P$ units) at $N=20$ 
$e$-folds before the end of inflation ($\phi_{N=20}$), at the inflationary exit ($\phi_{\rm end}$) and at 
the onset of the kinetic domination regime 
($\phi_{\rm kin}$).}\label{fig:chivsphi}
\end{figure}
The kinetic term can be made canonical by performing an additional field redefinition
\be
\frac{d\phi}{d\varphi}=k(\varphi) \,.
\ee
The relation between $\phi$ and $\chi$ is given by (see also Fig.~\ref{fig:chivsphi})
\begin{equation}\label{Wexact}
V=\left(\frac{\mu}{\chi}\right)^2=V_0 \left[\frac{\exp\left(-Y\right)}{Y}\right]^{2/\sigma}\,,
\end{equation}
with
\begin{equation}\label{Lamb}
Y=\frac{\sigma}{\kappa B(\chi)}=1+\frac12 \left[\frac{\phi^2}{\phi_t^2}+
\frac{\phi}{\phi_t }\sqrt{4+\frac{\phi^2}{\phi_t^2}}\right]\,.
\end{equation}
Here
\be\label{V0def}
V_0=\left(\frac\mu m\right)^2\left(\frac{\sigma}{\kappa} \right)^{2/\sigma}\,,
\ee
and 
\begin{equation}
\phi_t\equiv \frac{2M_P}{\sqrt{\kappa\sigma}}\,,
\end{equation}
denotes a \textit{transition} field value lying between the UV and IR fixed points. Indeed Eq.~\eqref{Wexact} implies
\begin{equation}
Ye^Y=\frac{\sigma}{\kappa}\left(\frac{\chi}{m}\right)^{\sigma} \,.
\end{equation}
This equation allows us to identify $Y$ with the Lambert function in Eq.~\eqref{Brun} and establishes the first equality 
in Eq.~\eqref{Lamb}. For the relation between $Y$ and $\phi$ we take into account that 
\begin{eqnarray}
\frac{d\phi}{dY}&=&\frac{d\phi}{d\varphi}\frac{d\varphi}{d\chi}\frac{d\chi}{dY} =\frac{M_P}{\chi}\sqrt{B}\frac{d\chi}{dY}
\\&=&\frac{\phi_t}{2}(Y^{-1/2}+Y^{-3/2})\,.
\end{eqnarray}
Integrating this expression we get the identity
\begin{equation}\label{phiW}
\frac{\phi}{\phi_t}=Y^{1/2}-Y^{-1/2} \,,
\end{equation}
which can be easily inverted to obtain the second equality in Eq.~\eqref{Lamb}. The relation 
between $\varphi$ and $\phi$ follows from $\varphi(\chi)$ in Eq.~\eqref{Wvarphi} and $\chi(Y(\phi))$ as 
given by Eqs.~\eqref{Wexact} and \eqref{Lamb}.

\begin{figure}
\centering
\includegraphics[scale=0.37]{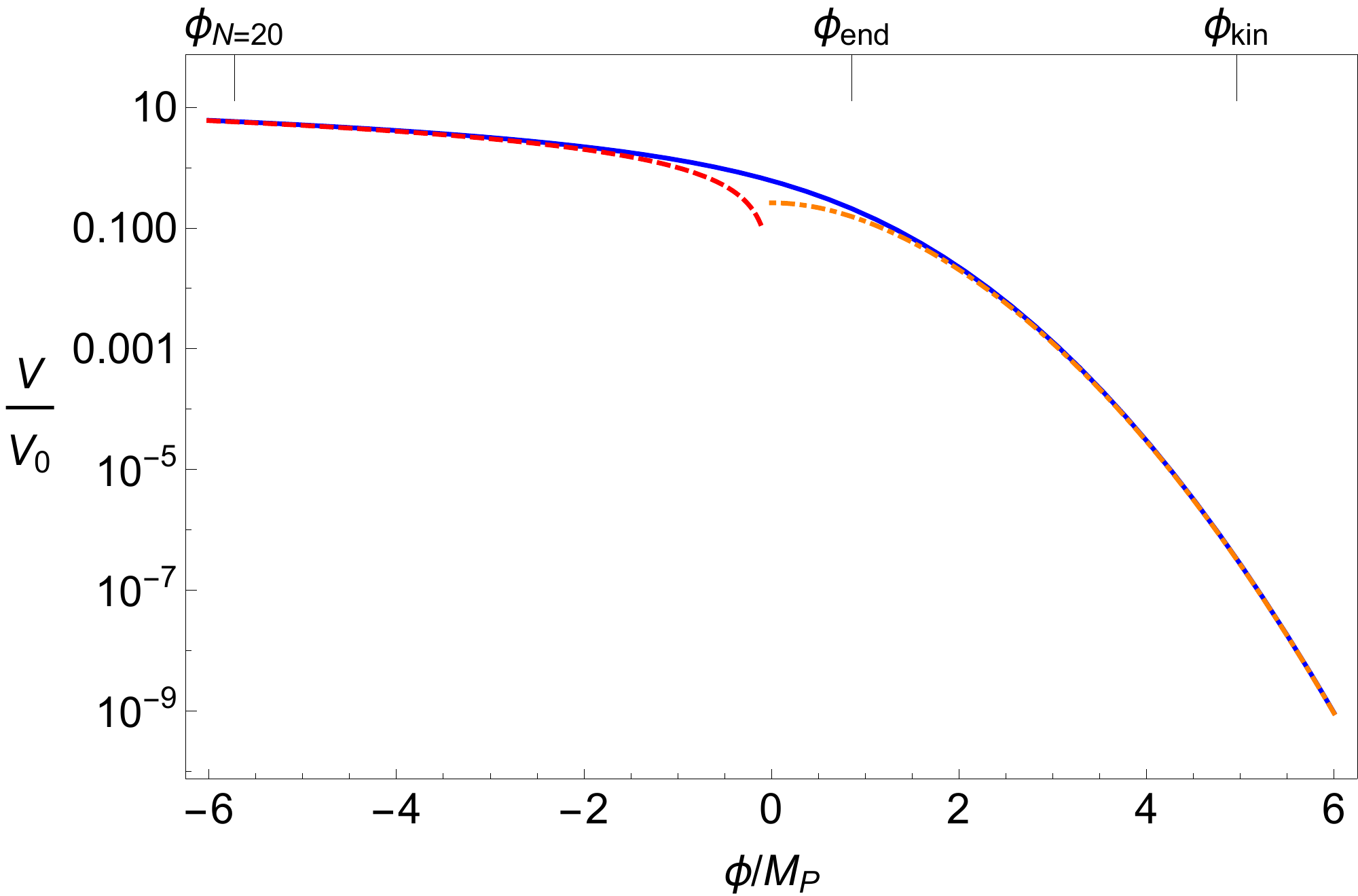}
\caption{The cosmon potential for $\sigma=4$ and $\kappa=1$. The blue line corresponds to the exact expression
\eqref{Wexact}, while the orange-dotted line
and  red-dashed lines are associated to the approximated expressions \eqref{Wsmall} and \eqref{Wlarge} 
respectively. For reference, we indicate the values of the cosmon field $\phi$ (in $M_P$ units) at $N=20$ 
$e$-folds before the end of inflation ($\phi_{N=20}$), at the inflationary exit ($\phi_{\rm end}$) and 
at the onset of the kinetic domination regime ($\phi_{\rm kin}$). }\label{fig:Wpot}
\end{figure}

In the canonical basis, the action takes the standard form
\ba\label{EFactionCN}
\frac{\cal L}{\sqrt{- g}} =
\frac{M_P^2}{2} R-\frac12 (\partial \phi)^2- M_P^4\,V(\phi)\,.
\ea
All dynamical information is now encoded in the effective potential $M_P^4 V(\phi)$, as given by 
Eqs.~\eqref{Wexact} and \eqref{Lamb}. The second term 
in Eq.~\eqref{Lamb} contains a linear piece in $\phi$. The potential $V(\phi)$ in Eq.~\eqref{Wexact} is 
therefore nonsymmetric for arbitrary $\sigma$. For $\phi\ll -\phi_t$ one gets $Y\approx \phi_t^2/\phi^2$ and Eq.~\eqref{Wexact} becomes
a power-law (chaotic) potential
\be\label{Wlarge}
V(\phi)\simeq  V_0\, \left(\frac{\phi^2}{\phi_t^2}\right)^{2/\sigma}= A\, \left(\frac{\phi^2}{M_P^2}\right)^{2/\sigma}\,,
\ee
with 
\be\label{Adef}
A\equiv \left(\frac\mu m\right)^2\left(\frac \sigma 2\right)^{4/\sigma}\,.
\ee
On the other hand, for $\phi \gg \phi_t$ one has $Y\approx \phi^2/\phi_t^2+2$ and $V(\phi)$ can be approximated by a 
mixed-quintessence potential
\be \label{Wsmall}
V(\phi) \simeq V_0\left[\frac{\exp\left({-\frac{\phi^2}{\phi_t^2}-2}\right)  }
{\frac{\phi^2}{\phi_t^2}+2}\right]^{2/\sigma}\,.
\ee
The comparison between the exact cosmon potential \eqref{Wexact} and the approximated expressions  
\eqref{Wlarge} and  \eqref{Wsmall} is shown in Fig.~\ref{fig:Wpot}. 

We recall that the ratio $\mu/m$ in Eq.~\eqref{Adef} is 
related to the integration constant $c_t$ determining the particular trajectory in the flow (cf. Eq.~\eqref{mdef}). Order 
one values of $c_t$ translate naturally into values of $A$ that are exponentially smaller than one, $A\sim \exp(-2 c_t)$. 
This provides for a natural explanation of the small amplitude of the primordial fluctuations. 

\section{Inflationary era} \label{sec:inflation}

We can now proceed to discuss the observable consequences of our model by using the standard methods developed 
for a canonically normalized  scalar field in the Einstein frame. If correctly defined and 
computed, the observable predictions cannot depend on the particular frame under consideration nor on the
precise scalar-field normalization. This is indeed verified by the following independent computations. These 
computations put our variable gravity framework in direct contact with the known properties of inflationary 
potentials and dynamical dark energy scenarios existing in the literature.

The approximate power-law form of the potential at $\phi\ll -\phi_t$ allows for inflation with the 
usual chaotic initial conditions. The Einstein-frame equation of motion for the cosmon field in a flat
Friedmann-Lema\^itre-Robertson-Walker Universe
\begin{equation}
ds^2=-dt^2+a^2(t) d{\bf x}^2\,,
\end{equation}
 is given by
\be\label{cosmoneq}
\ddot \phi+3H\dot\phi + M_P^4 \,V,_\phi=0\,,
\ee
with dots denoting derivatives with respect to the coordinate time $t$ and $H=\dot a(t)/a(t)$.
The Universe undergoes a phase of accelerated expansion if 
\be\label{eH}
\epsilon_H\equiv-\frac{\dot H}{H^2}< 1\,.
\ee
The evolution of the acceleration parameter \eqref{eH} can be determined by numerically solving 
Eq.~\eqref{cosmoneq} together with the Friedmann equations and standard slow-roll initial conditions. 
Depending on the value of $\phi_t$, the end of inflation for the inflationary potentials \eqref{Wexact} 
and \eqref{Wlarge} can take place at slightly different field values (the smaller the transition scale $\phi_t$, 
the smaller the difference). This change translates into a small variation in
the number of $e$-folds for the values of $\kappa$ and $\sigma$ we are interested in. Having this in mind, we
will estimate the inflationary observables using the simple power-law approximation \eqref{Wlarge}. 
\begin{figure}
\centering
\includegraphics[scale=0.37]{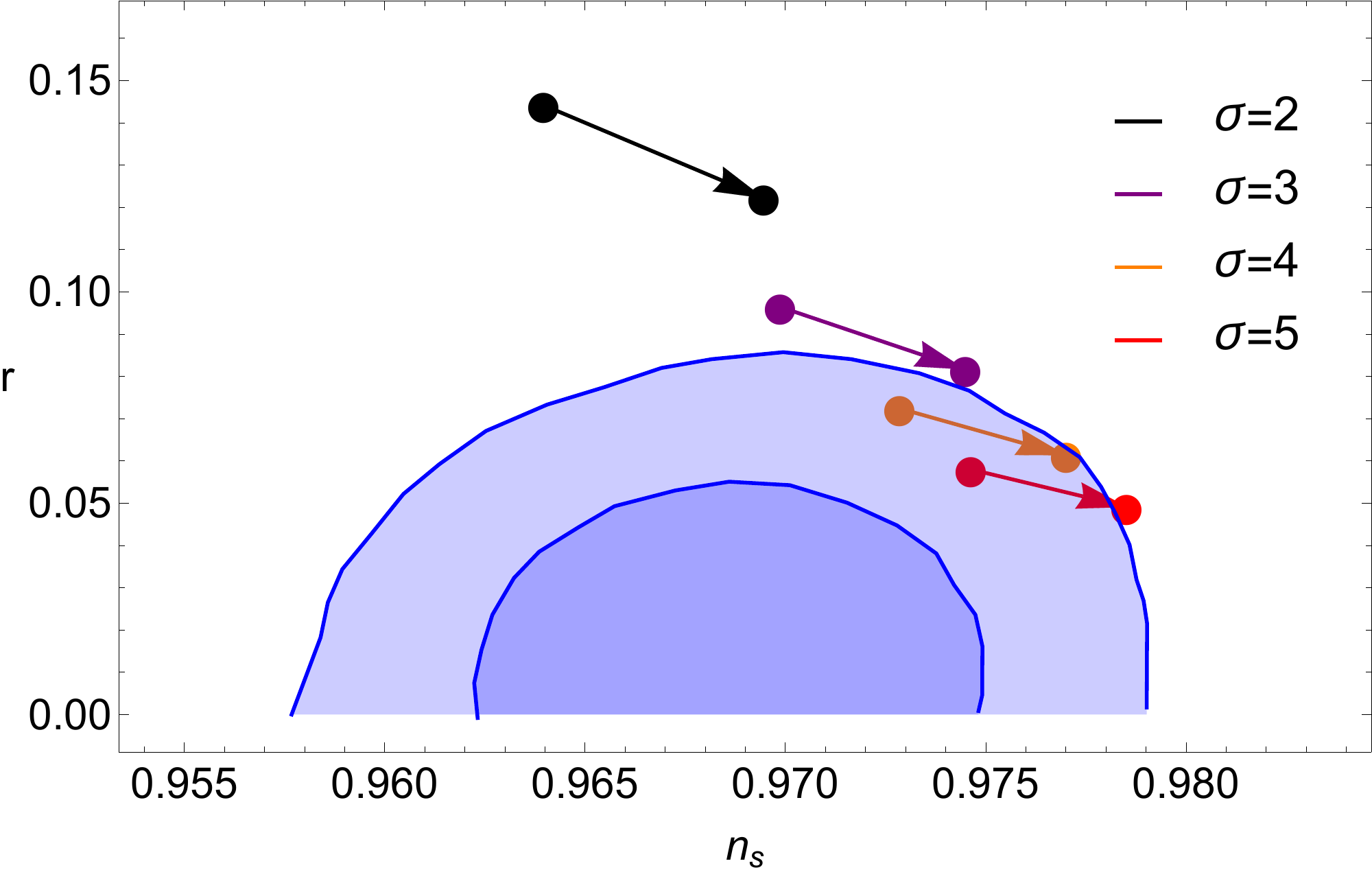}
\caption{Comparison between the inflationary predictions \eqref{nsr} and the latest Planck/BICEP2 data 
at 68\% and 95\% C.L. \cite{Ade:2015lrj,Ade:2015xua}.  The arrows go from the values obtained assuming $N=55$ $e$-folds 
to those for $N=65$ (cf. Section \ref{sec:heatingeff} for a more precise estimation of the number 
of $e$-folds). For $\sigma\gtrsim 4$, the predicted spectral tilt and tensor-to-scalar ratio lay 
inside the $2\sigma$ contour. 
} \label{fig:nsvsr}
\end{figure}
Following the standard  procedure, we obtain the following expressions for the spectral tilt and the tensor-to-scalar ratio 
\be\label{nsr}
1-n_s=\frac{2+\sigma}{\sigma N+ {1}} \,,\hspace{10mm} r=\frac{16}{\sigma N+{1}}\,,
\ee
with 
\be
N=\frac{\sigma}{8 M_P^2}\left[\left(\phi_{\rm hc}\right)^2-
\left(\phi_{\rm end}\right)^2\right]\,,
\ee
the number of $e$-folds between the horizon crossing of the relevant fluctuations  ($\phi=\phi_{\rm hc}$) and the end of 
inflation ($\phi_{\rm end}/M_P= 2\sqrt{2}/\sigma$). These Einstein-frame results improve the 
estimates in Ref.~\cite{Wetterich:2014gaa} by properly identifying the end of inflation with $\epsilon_H\approx \epsilon_V=1$.
\footnote{The estimates in Ref.~\cite{Wetterich:2014gaa} replace the 
denominators $\sigma N+1$ in Eq.~\eqref{nsr} by $\sigma N+3$, due to a small change in the precise definition of the end 
of inflation. While in the present work, the offset of inflation is defined to occur at $B_{\rm end}=2$, 
Ref.~\cite{Wetterich:2014gaa} takes $B_{\rm end}=6$.} The comparison of \eqref{nsr} with the latest 
cosmic microwave background (CMB) results is shown in Fig.~\ref{fig:nsvsr}. For an anomalous dimension $\sigma\gtrsim 4$, the 
inflationary predictions lay within the $2\sigma$ Planck/BICEP2 contour \cite{Ade:2015lrj,Ade:2015xua}.

The amplitude of scalar perturbations
\be\label{amplitude}
{\cal A}=\frac{V(\phi_{\rm hc})}{r M_P^4}=3.56\cdot 10^{-8},
\ee
together with Eqs.~\eqref{Wlarge} and \eqref{nsr} evaluated at horizon crossing  ($\phi=\phi_{\rm hc}$) determine 
the ratio
\ba\label{movermu}
\frac m\mu
&=&2^{\frac{1}{\sigma}-2}
\big(\sigma N+1)^{\frac12+\frac{1}{\sigma}}{\cal A}^{-\frac12}\,,
\ea 
and the associated trajectory in the flow.  For $\sigma=4$ and $N=60$, one obtains $m\simeq 10^{5}\mu$. Our 
results agree with Ref.~\cite{Wetterich:2014gaa}.

Once we have determined the Einstein-frame inflationary dynamics, we can always reinterpret our results in terms of 
the original variable gravity formulation. In particular, it is interesting to compare the value of the cosmon 
field $\chi$ during inflation with the scales $\mu$ and $m$. Combining Eqs.~\eqref{Wvarphi} and \eqref{amplitude} we get 
\be
\frac{\chi_{\rm hc}}{\mu}=\frac{1}{\sqrt{{\cal A}r}},
\ee
meaning that horizon crossing happens when $\chi\gg \mu$. This value is, however, much smaller than $m$, as can 
be easily seen by combining Eqs.~\eqref{Wlarge} and \eqref{Adef} and taking into account Eq.~\eqref{Wexact},
\be\label{chiphi}
\frac{\chi_{\rm}}{m}=
\left(\frac{4 M_P^2}{\sigma^2 \phi^2}\right)^{1/\sigma}\,.
\ee
 Evaluating the result at horizon crossing we get
\be
\frac{\chi_{\rm hc}}{m}=
\left(\frac{4 M_P^2}{\sigma^2 \phi_{\rm hc}^2}\right)^{1/\sigma}=\left(\frac{r}{32}\right)^{1/\sigma}\,,
\ee
with $r$ the tensor-to-scalar ratio in Eq.~\eqref{nsr}. The scale $m$ is indeed associated to the end of inflation   
\be
\frac{\chi_{\rm end}}{m}=\left(\frac{4 M_P^2}{\sigma^2 \phi_{\rm end}^2}\right)^{1/\sigma}=\frac{1}{2^{1/\sigma}}\,.
\ee

A simple overall picture of inflation arises. The inflationary phase corresponds to the vicinity of the UV fixed point 
for $\chi\lesssim m$. Close to a fixed point approximate scale symmetry is manifestly realized. This approximate symmetry 
is the origin of the almost scale invariant primordial fluctuation spectrum. For $\chi\approx m$ one observes the crossover 
from the vicinity of the UV fixed point to the vicinity of the IR fixed point. Scale invariance is substantially violated in 
this crossover region. This violation triggers the end of inflation. The scale $m$ is an 
integration constant of an almost logarithmic flow. Small values of $\mu/m$ arise therefore naturally, similarly to
the small ratio of the confinement scale in quantum chromodynamics as compared to some ``unification scale''. This provides 
for a small fluctuation amplitude 
\begin{equation}
{\cal A}=\frac{1}{32}\left[2(\sigma N+1)\right]^{1+\frac{2}{\sigma}}\frac{\mu^2}{m^2}\,,
\end{equation}
without the necessity for tuning.

\section{Kinetic dominated era} \label{sec:kinetial}

\begin{figure}
\centering
\includegraphics[scale=0.35]{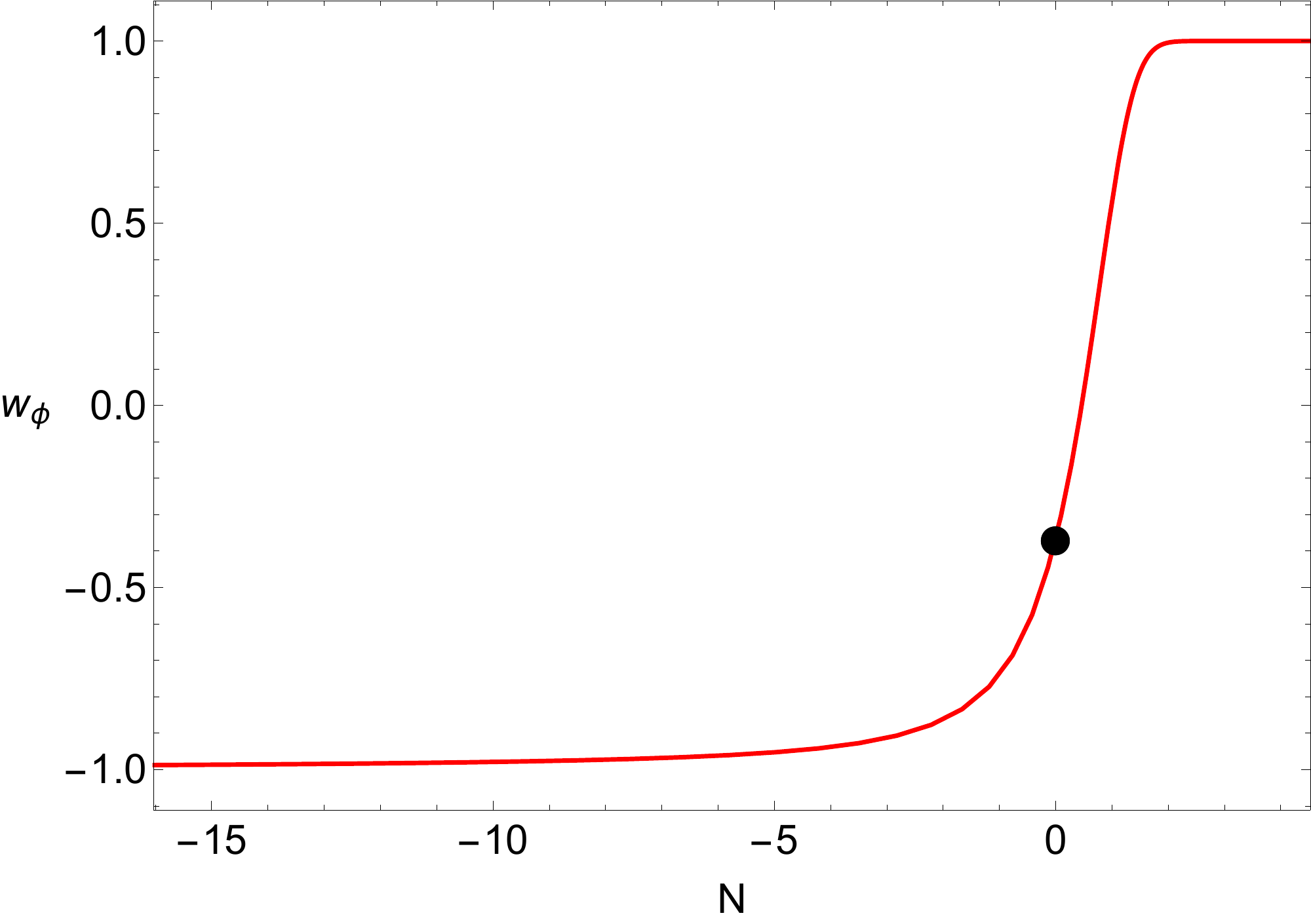}
\caption{The cosmon equation of state in the potential \eqref{Wexact} with $\sigma=4$ and $\kappa=1$ as 
a function of the number of $e$-folds $N$.  The end of inflation after $60$ $e$-folds is indicated with a black dot. The 
limit $w_\phi=1$ corresponds to a kinetic dominated regime.} \label{fig:eos}
\end{figure}

After inflation, the inflaton rolls down into the steep potential \eqref{Wsmall}, leading to a 
substantial decrease of the potential energy density. The evolution of the cosmon field becomes dominated by its
 kinetic energy and the heating of the Universe sets in. In this section, we consider
 the initial epoch in which the energy density into radiation is still small as compared to the energy density of 
 the cosmon. Such a period is usually referred as \textit{kination} or \textit{deflation} \cite{Spokoiny:1993kt}. 
 During this epoch, we can continue to use the cosmon-field equation \eqref{cosmoneq}. Neglecting the potential 
 energy density in the first approximation, the equation $\ddot \phi +3 H\dot \phi \simeq 0$ with  $H=1/(3t)$ admits a solution 
\begin{equation}\label{evolkin}
\phi(t)=\phi_{\rm kin}+\dot\phi_{\rm kin}t_{\rm kin}\log\left(\frac{t}{t_{\rm kin}}\right)\,,
\end{equation}
with $\phi_{\rm kin}$ and $\dot\phi_{\rm kin}$ the value of the field and its derivative at the onset of the 
kinetic era at $t_{\rm kin}$. During this regime the cosmon energy density 
scales as $a^{-6}$. This behavior is reflected by the cosmon equation of state
\be
w_\phi=\frac{p_\phi}{\rho_\phi}=\frac{\frac{1}{2}\dot \phi^2-M_P^4 V}{\frac{1}{2}\dot \phi^2+M_P^4 V}\approx 1\,.
\ee
In Fig.~\ref{fig:eos}, we show the numerical solution of the field equations for $w_\phi$ as a function of the number of 
$e$-folds $N$. The equation of state evolves rapidly toward $w_\phi=1$ after the end of inflation. 

For a realistic cosmology, the kinetic domination regime has to end before big bang nucleosynthesis. For 
this era to begin, the energy density of the cosmon field must be (dominantly) transmitted to the Standard Model degrees 
of freedom. In a nonoscillatory model like the one under consideration, a total decay of the 
inflaton field is not expected since $\dot\phi=0$ is not a solution of the equations of motion in the absence of 
a minimum. Also, highly effective processes such as parametric resonance cannot take place in a nonoscillatory model. 

Although an incomplete inflaton decay would constitute a serious drawback for most inflationary models, it does not
in the variable gravity scenario. To understand this, let us assume that a given heating mechanism is able to 
produce a partial depletion of the  cosmon condensate by the creation of relativistic particles. Even if the 
energy density of this component is initially 
very small, it will inevitably dominate the energy budget at later times. Indeed, during the kinetic dominated regime
the energy density  of the 
created particles scales as $\rho_{\rm r}\sim a^{-4}$, while that of the cosmon field evolves as $a^{-6}$.  The 
rapid decrease of the cosmon energy density will inevitably give rise to a late-time domination of the radiation 
component. 

The \textit{radiation temperature} $T_{\rm rad}$ at which the energy density of the created particles 
equals that of the cosmon ($\rho_{\rm r}^{\rm rad}=\rho_\phi^{\rm rad}$) can be defined as
\begin{equation}
T_{\rm rad}\equiv \left(\frac{30\,\rho_{\rm r}^{\rm rad}}{\pi^2 g^{\rm rad}_*}\right)^{1/4}\,,
\end{equation}
with  $g^{\rm rad}_*$ the effective number of relativistic degrees of freedom at that temperature. The quantity 
$T_{\rm rad}$ should be interpreted as the typical energy scale for the onset of radiation domination. It 
coincides with the heating 
temperature (usually called the reheating temperature) in the \textit{fast thermalization} limit.

For a simplified scenario, 
we may assume particle production to take place instantaneously at the onset of the kinetic regime. This motivates 
the introduction of a \textit{heating efficiency}, defined in this limit as 
\begin{equation}\label{thetadef}
\Theta\equiv \frac{\rho_{\rm r}^{\rm kin}}{\rho^{\rm kin}_\phi}\,.
\end{equation}
We will later extend the definition of $\Theta$ to smoother transitions. For the types of heating or entropy production 
mechanisms considered in this paper the parameter $\Theta$ is sufficient for a quantitative description of the cosmological history. 

For instant particle production the heating efficiency can be easily related to the radiation 
temperature $T_{\rm rad}$ by taking into account that 
\begin{equation}\label{fraction}
\Theta=\frac{\rho_{\rm r}^{\rm kin}}{\rho^{\rm kin}_\phi}=\left(\frac{a_{\rm kin}}{a_{\rm rad}}\right)^2 \,.
\end{equation}
We get
\begin{equation}\label{Treh}
T_{\rm rad}
= \left(\frac{30\,\Theta^3 \rho_\phi^{\rm kin}}{\pi^2 g_*^{\rm rad}}\right)^{\frac{1}{4}}
=\Theta^{\frac{1}{2}} \left(\frac{g_*^{\rm kin}}{g_*^{\rm rad}}\right)^{\frac{1}{4}}T_{\rm kin}\,,
\end{equation}
with 
\begin{equation}
T_{\rm kin}\equiv \left(\frac{30\,\rho_{\rm r}^{\rm kin}}{\pi^2 g^{\rm kin}_*}\right)^{1/4}\,,
\end{equation}
the temperature of the created particles at the onset of kinetic domination. 
\begin{table}
\begin{center}
\begin{tabular}{cccccc}    \toprule
$\sigma$ & $\phi_t/M_p$ & $H_{\rm end}$ (GeV) & $H_{\rm kin}$ (GeV) & 
\\\midrule
$2$   &$\sqrt{2}$ & $6.2 \times 10^{12}$ & $5.8\times 10^{10}$
\\
$3$ & $2/\sqrt{3}$  & $8.9 \times 10^{12}$ & $1.2\times 10^{11}$ 
\\
$4$ & 1 &$1.1 \times 10^{13}$ & $1.6 \times 10^{11}$
\\ 
\hline
\end{tabular}
\end{center}
 \caption{Approximate values of the Hubble rate at the end of inflation and at the onset of 
 the kinetic dominated era.  The numbers displayed were obtained by numerically solving the 
 equations of motion for $\kappa=1$ and different values of $\sigma$. The 
 end of inflation is defined by the condition $\epsilon_H=1$ with $\epsilon_H$ given by Eq.~\eqref{eH}. The beginning of the 
 kinetic era is defined by the time at which the effective equation-of-state parameter $w_\phi$ equals 1, up to one percent
 accuracy. 
 }\label{table0}
\end{table}
The longer is the kinetic regime, the 
smaller is the radiation temperature $T_{\rm rad}$. The heating efficiency \eqref{fraction} must be large enough to avoid conflicts 
with BBN. For the power-law inflationary potential \eqref{Wlarge}, the Hubble rate at the end of inflation/onset 
of kinetic domination is of order $10^{11}-10^{13}$ GeV (cf. Table \ref{table0}), with a slight dependence on 
the precise value of $\sigma$. Taking this into account, Eq.~\eqref{Treh} becomes
\begin{equation}\label{Tradnum}
\frac{T_{\rm rad}}{10^{14}\, {\rm GeV}}\simeq \, a\, 
 \Theta^{3/4} \left(\frac{H_{\rm kin}}{10^{11}}\right)^{1/2}  \,,
\end{equation}
with $a=8.65 (g_*^{\rm rad})^{-1/4}$.

The inflationary dynamics not only excites cosmon fluctuations but  also generates primordial gravitational 
waves (GW). In the 
postinflationary era, the amplitude of GW with superhorizon wavelengths remains constant 
until it reenters the horizon. When that happens, the logarithmic GW spectrum  scales as \cite{Sahni:1990tx}
\begin{equation}
 \Omega_{\rm GW}(k)=\frac{1}{\rho_c}\frac{d\rho_{\rm GW}}{d\ln k} \propto k^{2\left(\frac{3 w-1}{3w +1}\right)}\,,
\end{equation}
with $w$ the effective equation of state. For a radiation dominated expansion, the GW spectrum remains flat.  
However, for a kinetic dominated regime, the spectrum becomes blue tilted and may eventually dominate the total
energy budget.  

Nucleosynthesis constraints set an integral bound on the GW density fraction at BBN, namely \cite{Maggiore:1999vm}
\be\label{GWbound}
h^2\int^{k_{\rm end}}_{k_{\rm BBN}} \Omega_{\rm GW}(k) \,d\ln k\lesssim  10^{-5}\,,
\ee
with $h=0.678$ and $k_{\rm end}$ and $k_{\rm BBN}$ the momenta associated respectively to the horizon scale 
at the end of inflation and at BBN. The dominant contribution to this integral comes from momenta that left 
the horizon before 
the end of inflation and reentered during kinetic domination. For these modes 
($k_{\rm rad}<k<k_{\rm kin}$)~\cite{Giovannini:1999bh} (see also Refs.\cite{Sahni:1990tx,Giovannini:1998bp}),
\be\label{hbranch}
\Omega_{\rm GW}(k)=\varepsilon\, \Omega_{\gamma} h^2_{\rm GW}\left(\frac{k}{k_{\rm rad}}\right)
\ln^2\left(\frac{k}{k_{\rm kin}}\right)\,,
\ee
with 
\be
h_{\rm GW}^2=\frac{1}{8\pi}\left(\frac{H_{\rm kin}}{M_P}\right)^2
\ee
the dimensionless amplitude of gravitational waves. The present radiation content
in critical units $\rho_c=1.05\times 10^{-5} h^2 \,\textrm{GeV}\,\textrm{cm}^{-3}$ is given by
\be
\hspace{0mm}\Omega_\gamma \equiv\frac{\rho_{\gamma}(t_0)}{\rho_c(t_0)}=\hspace{-0.5mm}
2.6\times 10^{-5}h^{-2}\,.
\ee
 The factor 
\be
\varepsilon =\frac{81}{16\pi^3}\left(\frac{g_{\rm dec}}{g_{\rm th  }}\right)^{1/3}
\ee
takes into account the variation on the number of massless degrees of freedom between thermalization and decoupling 
\cite{Giovannini:1999bh}. For the Standard Model content ($g_{\rm th}=106.75$, $g_{\rm dec}=3.36$), we have 
$\varepsilon\simeq 0.05$.

Combining Eqs.~\eqref{GWbound} and \eqref{hbranch}, and neglecting a 
subleading logarithmic correction in the $k_{\rm kin}\gg k_{\rm rad}$ limit, we obtain 
\begin{eqnarray}\label{GWbound2}
2 \varepsilon h^2 \Omega_\gamma h_{\rm GW}^2 \left( \frac{k_{\rm kin}}{k_{\rm rad}}\right) \lesssim 10^{-5}\,.
\end{eqnarray}
 The ratio 
$k_{\rm kin}/k_{\rm rad}$ in this expression can be easily related to Eq.~\eqref{fraction} by taking into account that 
\begin{equation}\label{ratioks}
\frac{k_{\rm min}}{k_{\rm rad}}=\frac{a_{\rm kin}H_{\rm kin}}{a_{\rm rad}H_{\rm rad}}=
\frac{1}{\sqrt{2}}\left(\frac{\rho^{\rm kin}_\phi}{\rho_{\rm r}^{\rm rad}}\right)^{1/3}=\frac{1}{\sqrt{2}\Theta}\,.
\end{equation}
Using these results, the integral bound on the GW density fraction at BBN can be translated into a lower bound on 
the heating efficiency
\begin{eqnarray}\label{GWbound3}
\Theta \gtrsim \frac{10^5 \varepsilon \,h^2 \Omega_\gamma}{4\pi\sqrt{2}}  \left(\frac{H_{\rm kin}}{M_P}\right)^2\,.
\end{eqnarray}
For the typical values of $H_{\rm kin}$ in Table~\ref{table0} (and assuming $\varepsilon\simeq 0.05$), we get 
\begin{eqnarray}\label{GWbound4}
\Theta\gtrsim          10^{-17}\left(\frac{H_{\rm kin}}{10^{11}\,{\rm GeV}}\right)^2\,.
\end{eqnarray}
Using Eq.~\eqref{Tradnum}, this translates into a lower bound on the radiation temperature 
\begin{equation}
(g_*^{\rm rad})^{1/4}T_{\rm rad}\geq 225 \,{\rm GeV}\,. 
\end{equation}

\section{Heating }\label{sec:reheating}

Among the different heating mechanisms that have been proposed in the literature (see for instance 
Refs.~\cite{Peebles:1998qn,Hossain:2014xha,Chun:2009yu,Tashiro:2003qp,Sami:2004xk,Feng:2002nb,Dimopoulos:2002hm,
Liddle:2003zw,Sami:2003my}), there are two that 
can be naturally realized in a variable gravity framework: heating via gravitational interactions and heating via
matter couplings involving strong adiabaticity violations. 
In the following,  we will estimate the contribution of these heating scenarios to the heating efficiency \eqref{fraction} and 
the associated radiation temperature \eqref{Treh}. 

\subsection{Heating via gravitational interactions}

The simplest and most minimalistic heating mechanism is particle creation via gravitational 
interactions \cite{Spokoiny:1993kt,Ford:1986sy,Damour:1995pd}. \textit{Scalar} fields nonconformally coupled 
to the metric tensor are inevitably produced in an expanding background,\footnote{Note however that this 
mechanism does not apply to gauge bosons and chiral 
fermions since their evolution equations in a conformally flat geometry as Friedmann-Robertson-Walker 
are invariant under Weyl rescalings.} provided that they are light enough 
as compared to the Hubble rate. In our scenario, the scalar sector contains the Higgs doublet 
and the cosmon, but it can also include additional scalars as those appearing in extensions of the Standard Model such 
as grand unification. 

During a Hubble time, the gravitationally induced variation of the (relativistic) 
scalar energy density $\Delta \rho_{\rm r}\sim T_H^4$ is associated to an effective 
(Hawking) temperature $T_H=H/(2\pi)$. This effect has to compete with the dilution due to the 
expansion of the Universe, $\rho_{\rm r}\sim a^{-4}$. During kinetic domination $H\sim a^{-3}$ and $\Delta\rho_{\rm r}\sim a^{-12}$. In consequence, gravitational particle production is dominated by times close to the onset of the kinetic epoch, while later particle 
production becomes negligible. On the other hand, during the inflationary epoch $H$ is almost constant
and $\Delta \rho_{r}\sim \exp(-4 N)$, with $N$ the numbers of $e$-folds. Particle creation 
during the early stages of inflation is therefore exponentially diluted and can be also neglected. We conclude
that sizable entropy production due to gravitational interactions concerns only the epoch immediately after inflation. 

As seen in Fig.~\ref{fig:eos}, the kinetic dominated era starts soon after the end of inflation. The energy 
scale  of the relativistic scalars created at the onset of this regime is of order 
\begin{equation}
T_{\rm kin}= \delta\times  \frac{H_{\rm kin}}{2\pi}\,,
\end{equation}
with $H^2_{\rm kin}=\rho^{\rm kin}_p/(3M_P^2)$ and $\delta\sim {\cal O}(1)$ an efficiency parameter 
\cite{Spokoiny:1993kt,Ford:1986sy}. 
Taking this expression into account, Eq.~\eqref{fraction} becomes
\begin{equation}
\begin{split}\label{ratioGR}
\Theta&=\frac{\delta^4 \, g_*^{\rm kin} }{1440\pi^2} \left(\frac{H_{\rm kin}}{M_P}\right)^2 \\ &=10^{-19}\,
\delta^4 \, g_*^{\rm kin}\left(\frac{H_{\rm kin}}{10^{11}{\rm GeV}}\right)^2\,, 
\end{split}
\end{equation}
resulting in a radiation temperature
\begin{equation}
T_{\rm rad}=\frac{\delta^{3}}{24\pi^2}
\sqrt{\frac{g_*^{\rm kin}}{10}}\left(\frac{g_*^{\rm kin}}{g_*^{\rm rad}}\right)^{1/4}\frac{H^2_{\rm kin}}{M_P}\,,
\end{equation}
with $g_*^{\rm kin}$ the effective number of (scalar) relativistic degrees of freedom at the transition from inflation to 
the kinetic epoch. 

If the created scalar particles are allowed to interact after production via nongravitational 
interactions,\footnote{Note that this does not apply to gravitational waves, which cannot thermalize below 
the Planck scale \cite{Giovannini:1998bp}.} they will rapidly generate a thermalized plasma that should contain, 
at least, the Standard Model degrees of freedom. In that case, the radiation temperature $T_{\rm rad }$ can 
be associated to the heating temperature. The effects of partial thermalization can be incorporated into a 
modification of the efficiency parameter $\delta$.

Note that, although $T_{\rm rad}$ is typically above the BBN temperature $T_{\rm BBN}\simeq 0.5$ MeV, it is not 
high enough to satisfy the bound \eqref{GWbound4} for a moderate number of scalar fields. Indeed, 
combining Eqs.~\eqref{GWbound3} and \eqref{ratioGR} (and assuming $\varepsilon\simeq 0.05$) we get 
\begin{equation}\label{boundg}
\delta^4 g_*^{\rm kin} \gtrsim {\cal O} (10^{2})\,.
\end{equation}
Thus, even for ${\cal O}(1)$ efficiency, a large number of \textit{scalar} fields is 
required in order to satisfy the GW constraints. 

Independently of the plausibility of Eq.~\eqref{boundg}, gravitational particle production should not 
be considered a completely satisfactory heating mechanism. As argued in Ref.~\cite{Felder:1999pv}, the presence of light 
fields \textit{during inflation} could give rise to unwanted effects, such as the generation of secondary inflationary periods 
or the production of large isocurvature perturbations. As we will show in the next section, these problems, together with the inefficiency of gravitational particle production, can be easily solved in the presence of \textit{direct} couplings 
between the cosmon field and matter.

\subsection{Heating via matter interactions}\label{subsec:gravrh}

After Weyl rescaling the coefficients in the quadratic part of the effective action for matter fields 
generically depend on $\phi$ (see Ref.~\cite{Wetterich:1987fk} for the Higgs doublet). For a scalar field $h$, this dependence 
can be parametrized as 
\begin{equation}\label{HLag0}
\hspace{-2mm}\frac{{\cal L}_{\rm I}}{\sqrt{-g}} =-\frac{1}{2}\Big[(\partial h)^2 
+ \gamma(\phi)(\partial h^2 \,\partial \phi)  + M_h^2(\phi) h^2 \Big]. 
\end{equation}
The $\phi$-dependence of the effective action induces particle production if the 
coefficients $M_h^2(\phi)$ or $\gamma(\phi)$ change substantially with time.  Rapid variations of these 
functions are expected to occur during the crossover, where the dimensionless couplings and mass ratios of
matter fields must evolve from  their UV fixed-point values to those associated to the IR fixed
point.\footnote{In the Einstein frame, the matter fields must eventually decouple from the cosmon to avoid violations of 
the equivalence principle \cite{Uzan:2010pm, Wetterich:2003qb}. This is realized if an IR fixed point is approached.} 

To understand how a change in the effective couplings translates into particle production, let 
us consider the Einstein-frame equation of motion for the $h$ field 
\be\label{motionC}
(-\nabla^\mu \nabla_\mu + M^2_h(\phi)) h =\nu(\phi) h\,,
\ee
with 
\begin{equation}
\nu(\phi)=g^{\mu\nu} \nabla_\mu \left[\gamma(\phi)\partial_\nu\phi\right]\,.
\end{equation}
For $\gamma(\phi)\neq 0$ the field equation for $h$ contains derivative interactions. For 
the sake of simplicity, we will neglect this coupling in the following 
considerations and set $\gamma(\phi)=0$.\footnote{Derivative interactions give rise to similar particle production effects, see 
for instance Ref.~\cite{Lachapelle:2008sy}.} For a homogeneous cosmon field ($\phi=\phi(t)$), the mode equation
in Fourier space reads
\be\label{motion}
\ddot h_k+3 H \dot h_k+\left(\frac{k^2}{a^2}+M_h^2\right)h_k=0\,.
\ee
The friction term in this expression can be eliminated by performing a field redefinition 
$h\rightarrow a^{-3/2} h$. Doing this, we get a time-dependent harmonic oscillator equation
\begin{equation}\label{perteq}
\ddot h_k+\omega_k^2(t)h_k=0\,,
\end{equation}
with
\begin{equation}
\omega_k^2(t)=\frac{k^2}{a(t)^2}+M^2_h(t) +\Delta_a\,,
\end{equation}
and
\ba\label{Deltaa}
&&\Delta_a =-\frac{3}{4}\frac{\dot a^2}{a^2}-\frac{3}{2}\frac{\ddot a}{a}\,.
\ea

The term $\Delta_a$ is responsible for the gravitational particle production 
discussed in Section \eqref{subsec:gravrh}. In the 
presence of direct couplings between the inflaton and matter fields this term is expected to be subdominant 
and it will be neglected in what follows.  

The solutions of the mode equation \eqref{perteq} could be used to compute the propagator for the $h$-field in 
the time-dependent background $\phi(t)$, along the lines of Ref.~\cite{Wetterich:2015gya}. From this, particle
creation can be directly extracted. 
We will follow here the more conventional approach based on the operator formalism.

Let us describe the solutions of the mode equation \eqref{perteq} in terms of
positive- and negative-frequency adiabatic solutions $\sim \exp(\pm  i \int_0^t  dt' \omega_k(t'))$, namely
\begin{equation}\label{ukplane}
h_k(t)=\frac{1}{\sqrt{2\omega_k}}\left[A_k(t)+ B_k(t)\right],
\end{equation}
with 
 \begin{eqnarray}
A_k(t)&\equiv& \alpha_k(t) e^{-i\int_0^t d t'\omega_k(t')}\,,\label{Ak} \\ 
B_k(t)&\equiv& \beta_k(t) e^{i\int_0^t d t'\omega_k(t')}\,. \label{Bk}
\end{eqnarray}
The time dependence of the functions $\alpha_k(t)$ and $\beta_k(t)$ has to ensure that the mode 
equation \eqref{perteq} is obeyed. We will require $\alpha_k(t)$ and $\beta_k(t)$ to satisfy the differential equations
\begin{eqnarray}
\dot \alpha_k(t)&=&\frac{\dot \omega_k}{2\omega_k} e^{2i\int_0^t d t'\omega_k(t')}\beta_k(t)\,,\label{alphaevol}\\
\dot \beta_k(t)&=&\frac{\dot \omega_k}{2\omega_k} e^{-2i\int_0^t d t'\omega_k(t')}\alpha_k(t)\,.\label{betaevol}
\end{eqnarray}
These conditions induce the following evolution equations for $A_k(t)$ and $B_k(t)$  
\begin{eqnarray}\label{ABeq}
\dot{A}_k(t)&=& \frac{\dot{\omega}_k}{2\omega_k}B_k(t) - i \omega_k A_k(t)\,, \\
\dot{B}_k(t)&=& \frac{\dot{\omega}_k}{2\omega_k}A_k(t) + i \omega_k B_k(t)\,.  \label{ABeq2}
\end{eqnarray}
The insertion of these equations into Eq.~\eqref{ukplane} yields indeed the mode equation \eqref{perteq}.

By virtue of Eqs.~\eqref{alphaevol} and \eqref{betaevol} one obtains the conservation equation 
\begin{equation}
\partial_t(\vert\alpha_k(t)\vert^2-\vert\beta_k(t)\vert^2)=0\,. 
\end{equation}
In particular, the Wronskian condition 
\begin{equation}\label{Wronskian}
\lvert \alpha_k(t) \rvert^2 - \lvert \beta_k(t) \rvert^2 = 1\,, 
\end{equation}
is preserved in time.  The condition \eqref{Wronskian} arises in the operator formalism from the commutation 
relations of creation and annihilation operators for free fields. Extracting the propagator as the inverse of the second 
functional derivative of the effective action, this condition is induced by the inhomogeneous term in the propagator equation \cite{Wetterich:2015gya}.

The occupation number of particles at time $t$ can be identified with 
\begin{equation}\label{Nk}
n_k=\vert B_k(t)\vert^2\,. 
\end{equation}
The number and energy density of the created particles is therefore given by
\begin{eqnarray}
n_h(t)&=& \frac{1}{a^3(t)}\int\frac{d^3k}{(2\pi)^3}n_k\,, \\ 
\rho_h(t)&=& \frac{1}{a^3(t)}\int\frac{d^3k}{(2\pi)^3} \omega_k n_k\,.
\end{eqnarray}
Using
\begin{equation}
\dot h_k(t)=-i\sqrt{\frac{\omega_k}{2}}(A_k(t)-B_k(t))\,,
\end{equation}
one infers the relation
\begin{equation}
\vert A_k(t)\vert^2+\vert B_k(t)\vert^2=
\frac{1}{\omega_k}\vert \dot h_k(t)\vert^2+\omega_k\vert h_k(t)\vert^2\,.
\end{equation}
This, together with the condition $\vert A_k\vert^2-\vert B_k\vert^2=1$, translates for the occupation numbers to
\begin{equation}
n_k=\frac{1}{2\omega_k}\left(\vert\dot h_k\vert^2+\omega_k^2\vert h_k\vert^2\right)-\frac{1}{2}\,. 
\end{equation}

Vacuum initial conditions correspond to $\alpha_k(t_{\rm init}) = 1$ and $\beta_k(t_{\rm init}) = 0$ and 
therefore to $n_k(t_{\rm init})=0$. In terms of  
$h_k$ these initial conditions become
\begin{eqnarray}
h_k(t\rightarrow t_{\rm init})&=&\frac{1}{\sqrt{2\omega_k(t)}} e^{-i\int_0^t d t'\omega_k(t')} \,. 
\end{eqnarray}

For particle production to be efficient the adiabaticity 
condition $\vert \dot \omega_k\vert \ll \omega_k^2$ must be significantly violated~\cite{Kofman:1997yn}, as clearly visible 
in Eq.~\eqref{ABeq2}. The energy density of the produced particles obeys
\begin{equation}
\left(\partial_t+3 H\right)\rho_h =\frac{1}{2a^3}\int \frac{d^3k}{(2\pi^3)}\dot\omega_k 
\left(2\omega_k\vert h_k\vert^2-1\right)\,. \nonumber
\end{equation}
The production term in the right-hand side is indeed proportional to $\dot \omega_k$. It has to compete with 
the Hubble damping in the left-hand side.

In the absence of derivative interactions (i.e. for $\gamma(\phi)=0$ in Eq.~\eqref{HLag0}), the 
cosmon field $\phi$ couples to the matter field $h$ only through the effective mass function $M_h^2(\phi)$. In a 
natural and phenomenologically successful scenario, this function should satisfy the following criteria:
\begin{enumerate}[i)]
 \item It should be large enough during inflation to retain the single-field inflationary picture and avoid the 
generation of large isocurvature perturbations.
\item It should rapidly vary at the end of inflation ($\phi\gtrsim \phi_{\rm end}$) to heat the Universe 
via violations of the adiabaticity condition $\vert \dot \omega_k\vert \ll \omega_k^2$.
\item It should eventually become independent of $\phi$ if the field $h$ is the Higgs doublet. This reflects 
scale symmetry in the Standard Model sector as required by the bounds on the variation of the ratio of the Fermi scale over 
the Planck scale since nucleosynthesis.
\end{enumerate}

We assume here that the crossover from the UV to the IR fixed point, which is reflected in the change of the cosmon kinetic 
term and associated to the end of inflation, leaves also its traces in the matter sector. In the field range $\chi\approx m$, 
the couplings of $h$ to $\chi$, and correspondingly to $\phi$, are therefore expected to undergo significant changes. This 
provides for a natural scenario where $M_h^2(\phi)$ can change rapidly from large to small values at the end of inflation. 
The conditions i)-iii) can be viewed as the imprint of the crossover in the matter sector.
\subsection{Workout example}

A possible parametrization of $M_h^2(\phi)$ satisfying the above requirements is $M_h^2(\phi)\equiv \epsilon(\phi) M_P^2$ with 
\be\label{ephi2}
\epsilon(\phi)=
 \epsilon_\infty+ \epsilon_1 \,\left[\frac{\exp\left(-Y_\epsilon(\phi)\right)}{Y_\epsilon(\phi)}\right]^{\sigma_h/2}\,.
\ee
Here
\be\label{Lambe}
Y_\epsilon(\phi)=1+\frac12 \left[\frac{\phi^2}{\phi_\epsilon^2}+
\frac{\phi}{\phi_\epsilon}\sqrt{4+\frac{\phi^2}{\phi_\epsilon^2}}\right]\,,
\ee 
and $\sigma_h$, $\epsilon_\infty$, $\epsilon_1$ and $\phi_\epsilon$ are taken to be positive constants. The shape of 
$\epsilon(\phi)-\epsilon_\infty$ mimics the form of the cosmon potential \eqref{Wexact}, with $\sigma_h$, $\phi_\epsilon$ and the amplitude $\epsilon_1$ left free. The detailed structure of this parametrization is chosen for illustration 
purposes only. Alternative choices sharing the features described in i), ii) and iii) could be used without
modifying the conclusions below.
\begin{figure}
\centering
\includegraphics[scale=0.8]{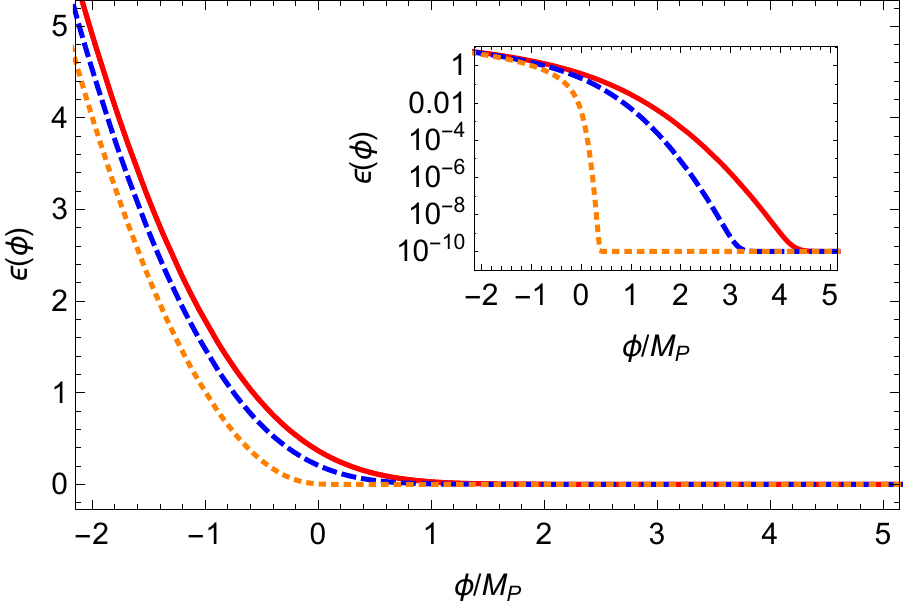}
\caption{The effective coupling $\epsilon(\phi)$ for $\sigma_h=2$, $\epsilon_\infty=10^{-10}$, $\epsilon_1 M_P^2= \phi_\epsilon^2$ and
different values of $\phi_\epsilon$. All cases share the same asymptotic behavior during inflation. The red-solid, 
blue-dashed and orange-dotted lines correspond respectively  to $\phi_\epsilon=M_P,\,0.75 \,M_P$ and $10^{-1}\,M_P$. In 
the inset we plot $\epsilon(\phi)$ logarithmically in order to better resolve the approach to zero and to 
facilitate the comparison with Fig.~\ref{fig:Wpot}.} \label{fig:Lamb}
\end{figure}

The behavior of Eq.~\eqref{ephi2} for different values of $\phi_\epsilon$ is shown in Fig.~\ref{fig:Lamb}. It 
describes the evolution from a UV fixed point\footnote{In the far UV ($\phi\gg\phi_t,\phi_\epsilon$), 
this corresponds in the \textit{scaling-frame} to a flow equation
\begin{equation}
\mu\partial_\mu \epsilon(\chi)\approx \bar\sigma_h \epsilon(\chi) \,,
\end{equation}
with $\bar \sigma_h=\sigma_h\sigma/2$.}
\begin{equation}\label{floweps}
\phi \,\partial_\phi \epsilon (\phi)\approx \sigma_h \epsilon(\phi)(1-\phi^2_\epsilon/\phi^2) \,, 
\end{equation}
 approached for $\phi \rightarrow -\infty$ with anomalous dimension $\sigma_h$, to an IR fixed point where 
 $\epsilon\approx \epsilon_\infty$. The constant $\phi_\epsilon$ encodes the location of the transition. The 
 smaller the value of $\phi_\epsilon$, the longer the effective coupling stays in the vicinity of the UV fixed point. 
 
 For large values of $\phi_\epsilon$, we can make use of Eq.~\eqref{chiphi} to relate this parameter to 
 the crossover scale $m$ signaling the end of inflation
\be
\frac{\chi_{\rm \epsilon}}{m}=\left(\frac{\kappa}{\sigma}\frac{\phi^2_t}{ \phi_{\epsilon}^2}\right)^{1/\sigma}\,.
\ee 
Values of $\phi_\epsilon^2$ larger than $\phi_t^2$ correspond to values of $\chi_\epsilon$ 
smaller than the crossover scale $m$. For $\phi_\epsilon/\phi_t\ll 1$, the transition in 
$M_{h}(\phi)$ occurs within a short period before the end of inflation.  In this region, the size of 
$\phi_\epsilon$ mainly determines the sharpness of the crossover, with small $\phi_\epsilon$ leading to a 
more abrupt transition. We could introduce and additional parameter $\phi_l$ for the \textit{timing} of the 
transition, e.g. by replacing $\phi$ in Eq.~\eqref{ephi2} by $\phi-\phi_l$. All our models can account for a gauge 
hierarchy if $\epsilon_\infty \ll 1$ with $\epsilon\gtrsim 1$ for $\phi\rightarrow -\infty$ ($\chi\rightarrow 0$) 
and $\epsilon\rightarrow \epsilon_\infty$ for $\phi\rightarrow \infty$ ($\chi\rightarrow \infty$).

The occupation numbers and the energy density of created particles for a given set of
parameters $(\sigma_h,\epsilon_\infty,\epsilon_1,\phi_\epsilon)$ can be computed by numerically solving 
the mode equation \eqref{perteq} with vacuum initial conditions. Let us consider for concreteness an 
anomalous dimension $\sigma_h=2$ in an inflationary model with $\sigma=4$ and $\kappa=1$. For this 
choice of parameters,  the interaction Lagrangian \textit{during inflation} ($\phi\ll -\phi_\epsilon$) 
contains a quartic term $-(1/2)\,g^2\, \phi^2 h^2$
with $g^2 \equiv \epsilon_1 M_P^2/\phi_{\epsilon}^2$. To retain the predictions of single-field inflation, we
will require the effective mass $g\vert \phi\vert$ of the scalar field $h$ during inflation to be larger than the mass of 
 the cosmon. In the leading order approximation \eqref{Wlarge} the squared cosmon mass 
 $M_c^2\equiv M_P^4\, \partial^2 V /\partial \phi^2$ reads 
\begin{equation}\label{MCapprox}
M_c^2=\frac{4A(4-\sigma)}{\sigma^2}
\left(\frac{\phi^2}{M_P^2}\right)^{\frac{2}{\sigma}-1} M_P^2\,.
\end{equation}
This expression vanishes for $\sigma=4$ and is generically suppressed by the small factor 
$A$, cf. Eqs.~\eqref{Adef} and \eqref{movermu}. Unless $g$ is tiny, the effective mass of the $h$ field during 
inflation will be significantly larger than the cosmon mass. For our practical example we choose $g^2=0.1$, while 
keeping $\phi_\epsilon$ as a free parameter. For $\epsilon_\infty$ 
 we take $\epsilon_\infty=10^{-10}$, which translates into an asymptotic $h$ mass of order 
 $\epsilon_\infty^{1/2} M_P \simeq {\cal O}(10^{13}$ GeV) at $\phi\rightarrow \infty$. Smaller values of 
 $\epsilon_\infty$, as those required if $h$ is the Higgs doublet, will not change our discussion.
 
\begin{figure}
\centering
\includegraphics[scale=0.38]{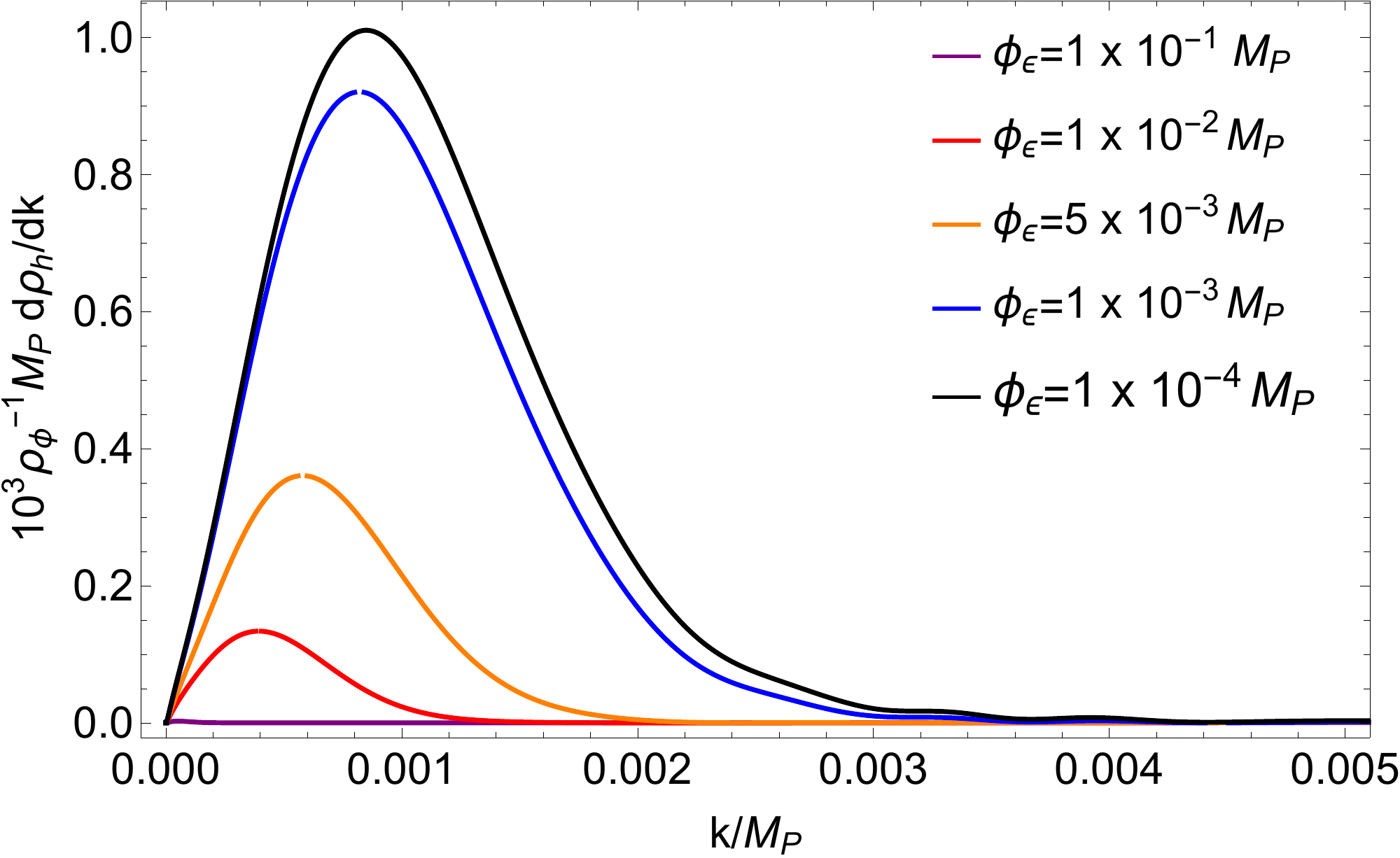}
\caption{The spectra of produced $h$ particles for  
$\sigma_h=2$, $\epsilon_\infty=10^{-10}$, $\epsilon_1=10^{-1}\phi_\epsilon^2/M_P^2$ and different 
choices of $\phi_\epsilon$. All cases are evaluated at the onset of the kinetic
domination regime. At that time, $H_{\rm kin}\simeq 1.63 \times 10^{11}\, {\rm GeV}$ and
$\rho^{\rm kin}_{\phi}\simeq (8.4 \times 10^{14}\, {\rm GeV})^4 $.}\label{fig:spectra}
\end{figure}

 The numerical results for  different values 
 of $\phi_\epsilon$ are summarized in Fig.~\ref{fig:spectra} and Table \ref{table1}. All cases are evaluated at the onset of 
 the kinetic domination regime.  At that time, $H_{\rm kin}\simeq 1.63 \times 10^{11}\, {\rm GeV}$ and
 $\rho^{\rm kin}_{\phi}\simeq (8.4 \times 10^{14}\, {\rm GeV})^4 $. In agreement with the analytical 
 estimates presented in Appendix \ref{sec:estimates}, smaller choices of $\phi_\epsilon$ translate into 
 more significant particle production.

The energy density of the produced particles can be compared with the energy density of the cosmon at the onset of
radiation domination. The resulting heating efficiency \eqref{fraction} determines the radiation temperature $T_{\rm rad}$ via 
Eq.~\eqref{Treh}. As shown in Table \ref{table1}, this temperature is well above the BBN temperature
($T_{\rm BBN}\simeq 0.5$  MeV) and 
significantly exceeds the energy scale associated to gravitational particle 
production. Note also that the GW bound \eqref{GWbound4} can be easily 
satisfied even if only one matter field violates the adiabaticity condition. If this 
condition is violated in  $n_a$ channels,  the ratio \eqref{fraction} and the associated radiation 
temperature are enhanced by  a factor $\Theta\rightarrow n_a \Theta$ and $T_{\rm rad}\rightarrow \sqrt{n_a} T_{\rm rad}$.
Since the main aspects of particle creation due to a \textit{single} violation of adiabaticity are independent of the spin 
of the particle, the above results can be extended to fermionic species \cite{Greene:1998nh,Greene:2000ew}.

\begin{table}
\begin{center}
\begin{tabular}{cccc}    \toprule
$\phi_\epsilon/M_P$ &  $\Theta $ & $(g_*^{\rm rad})^{\frac{1}{4}}T_{\rm rad}$ (GeV)
 \\\midrule
$1\times 10^{-1}$  & $ 2.5 \times 10^{-9}$   & $3.9\times 10^{8}$  \\ 
$1\times 10^{-2}$  & $ 9.7 \times 10^{-8}$   & $6.1\times 10^{9}$  \\ 
$5\times 10^{-3}$  & $ 3.3 \times 10^{-7}$   & $1.5\times 10^{10}$  \\ 
$1\times 10^{-3}$  &$1.2\times10^{-6}$ & $ 4.0\times 10^{10}$ \\
$1\times 10^{-4}$   & $1.4\times 10^{-6}$   &$4.4\times10^{10}$ \\
\hline
\end{tabular}
\end{center}
 \caption{Heating efficiency $\Theta$, as defined by the ratio  \eqref{fraction} between the radiation 
 and cosmon energy densities at the onset of kinetic 
 domination, and the associated radiation temperature \eqref{Treh} for different values of $\phi_\epsilon$. Parameters are
 taken as $\sigma_h=2$, $\epsilon_\infty=10^{-10}$ and $\epsilon_1=10^{-1}\phi_\epsilon^2/M_P^2$. At the onset of 
 kinetic domination one has $H_{\rm kin}\simeq 1.63 \times 10^{11}\, {\rm GeV}$ and 
 $\rho^{\rm kin}_{\phi}\simeq (8.4 \times 10^{14}\, {\rm GeV})^4 $.}\label{table1}
\end{table}

We finish this section by noticing that the heating scenario presented here is 
conceptually different from the \textit{instant ``reheating''} mechanisms 
\cite{Felder:1999pv,Felder:1998vq} 
appearing in most quintessential inflation models \cite{Hossain:2014xha,Tashiro:2003qp,Sami:2004xk}. \textit{Instant 
``reheating''} is usually formulated in terms of $\phi$-symmetric interactions among the inflaton field $\phi$ and
some scalar particle $X$ which is itself coupled to fermions via Yukawa interactions
\begin{equation}\label{Linstant}
\frac{{\cal L}_{\rm I}}{\sqrt{-g}}=-\frac{1}{2}g^2\phi^2 X^2 -y_\psi X\bar \psi \psi \,.
\end{equation}
As in our case, a small fraction of $X$ particles is automatically generated at the end of inflation via the
violation of the adiabaticity condition at $\phi\approx 0$. After particle production, the inflaton field rolls 
down the quintessence potential \eqref{Wexact}. This rolling increases the effective mass of the $X$ 
field ($m_X(\phi)= g \vert \phi\vert$) and amplifies its energy density and the 
probability to decay into fermions ($\Gamma_{X\rightarrow\bar\psi\psi} \propto m_X(\phi)$). As argued in 
Ref.~\cite{Felder:1999pv}, to avoid significant backreaction effects  into the inflaton evolution equation
\begin{equation}
\ddot \phi+3 H\dot \phi +V_{,\phi}=-g^2\phi \langle X\rangle^2\,,
\end{equation}
 the decay into fermions should take place soon after particle production. This requirement translates into a 
 mild condition relating the couplings $y_t$ and $g^2$ \cite{Felder:1999pv}.

Although \textit{instant ``reheating''} is a highly 
efficient mechanism that could give rise to radiation temperatures well above those displayed in Table~\ref{table1}, we 
find it difficult to implement in a variable gravity scenario like the one under consideration. For a simple crossover 
a monotonic dependence of $M_h^2(\phi)$ on $\phi$ seems more natural. We emphasize that an ``\textit{instant feeding}'' 
of the created particles is not necessary in our scenario, If the fraction of 
energy depleted out of the cosmon component exceeds the GW bound \eqref{GWbound4}, the Universe will become safely 
dominated by radiation before BBN. Our scenario does not suffer from backreaction problems since 
the mass of the $h$ field is a monotonically decreasing function of the inflaton field $\phi$.

\section{Hot Big Bang era}\label{sec:hbb}

In the Einstein frame, the evolution during the hot big bang era can be described in terms of the Friedmann equations and the Klein-Gordon equation for the cosmon field 
\begin{eqnarray}
&&H^2=\frac{1}{3M_P^2}\left(\rho_\phi+\rho_R+\rho_M\right)\,, \label{ME1} \\
&& \dot H=-\frac{1}{2M_P^2}\left(\dot\phi^2+\frac{4}{3}\rho_R+\rho_M\right)\label{ME1b} \,,\\
& &\ddot{\phi}+3H \dot{\phi}+M_P^4 V_{,\phi}=0\,,\label{ME2} 
\end{eqnarray}
with $\rho_\phi=\dot\phi^2/2+M_P^4 V$ and  $\rho_R$ and $\rho_M$ the energy densities of radiation and nonrelativistic 
matter with conservation equations  
\begin{eqnarray}
\dot{\rho}_R+4H \rho_R=0\,,\hspace{5mm} \dot{\rho}_M+3H \rho_M=0\,.\label{ME3} 
\end{eqnarray}

\subsection{Onset of radiation domination}

The evolution of  the dark energy equation of state $w_\phi=p_\phi/\rho_\phi$ and the (normalized) energy densities 
$\Omega_i\equiv \rho_i/(3M_P^2H^2)$ with $i=\phi, R$ and $M$ can be obtained by numerically solving the cosmological 
equations \eqref{ME1}-\eqref{ME3}. We assume a 
number of production channels of the order of the number 
of degrees of freedom in the Standard Model, ${\cal O}(10^2)$. The results for $\sigma=4$, 
$\kappa=1$ and $\Theta= 10^{-4}$ are shown in Figs.~\ref{fig:wphi},\,  
\ref{fig:omegas} and \ref{fig:omegadetail}.  
The qualitative behavior of the observables in these figures can be understood as follows:
\begin{figure}
\centering
\includegraphics[scale=0.37]{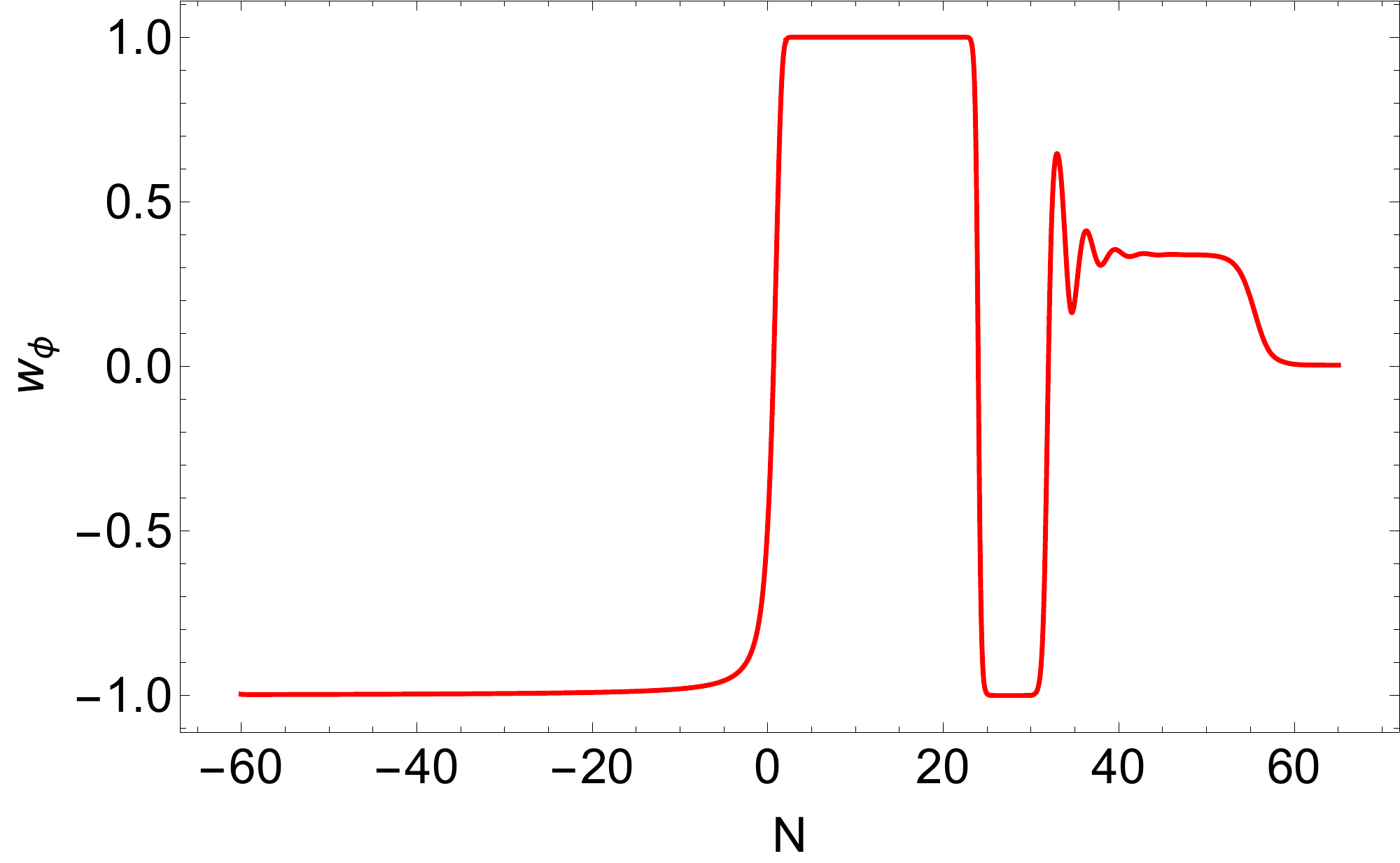}
\caption{Evolution of the dark energy equation of state $w_\phi=p_\phi/\rho_\phi$ from the inflationary 
era to the matter dominated era as a function of the number of $e$-folds. The 
end of inflation corresponds to $N=0$. For this plot, we chose $\sigma=4$, $\kappa=1$ and assumed a 
heating efficiency $\Theta= 10^{-4}$.} \label{fig:wphi}
\end{figure}
\begin{figure}
\centering
\includegraphics[scale=0.37]{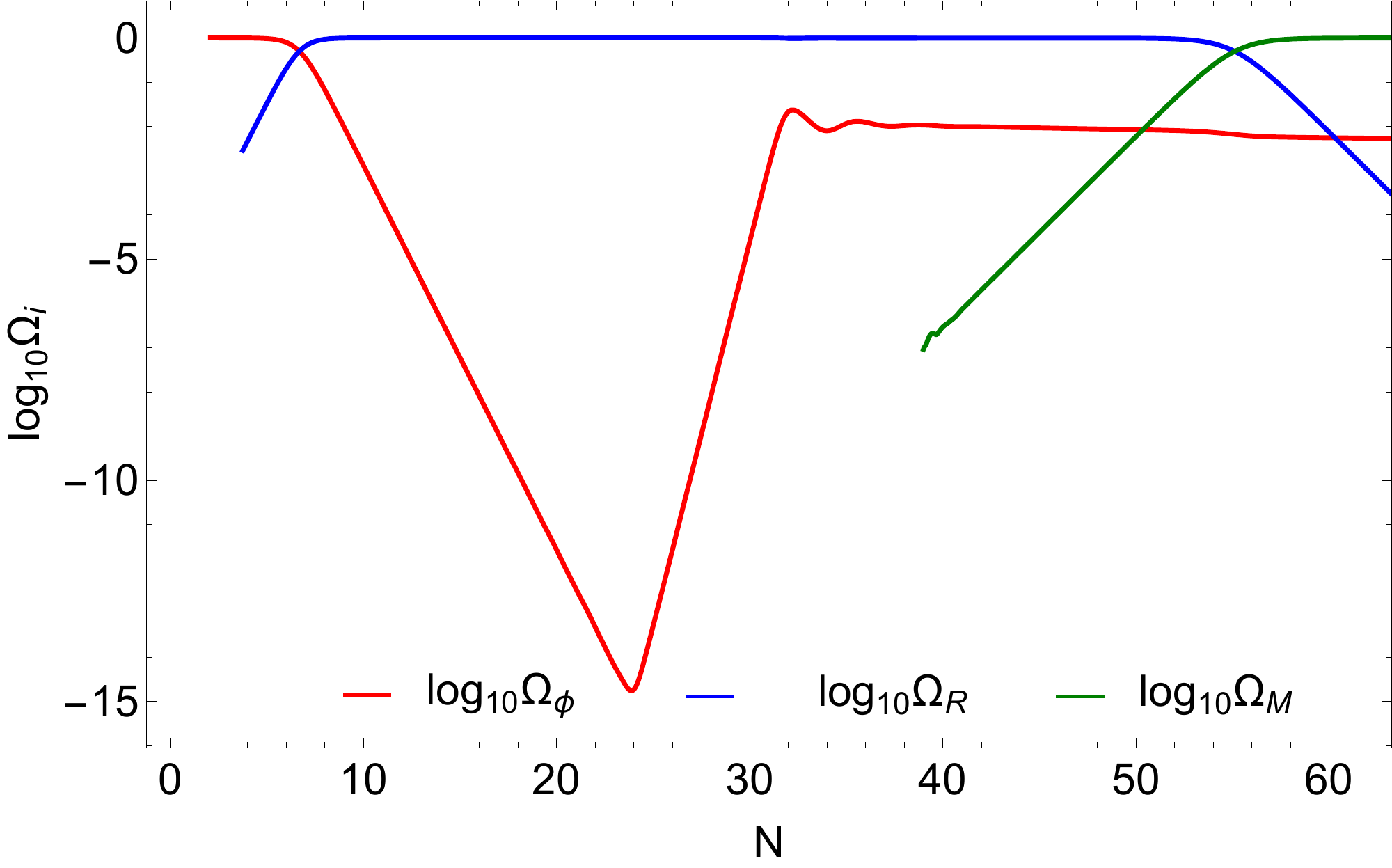}
\caption{Postinflationary evolution of the density parameters $\Omega_\phi$, $\Omega_R$ and $\Omega_M$  as 
a function of the number of $e$-folds. The end of inflation corresponds to $N=0$. Parameters are chosen as 
in Fig.~\ref{fig:wphi}.} \label{fig:omegas}
\end{figure}
\begin{figure}
\centering
\includegraphics[scale=0.37]{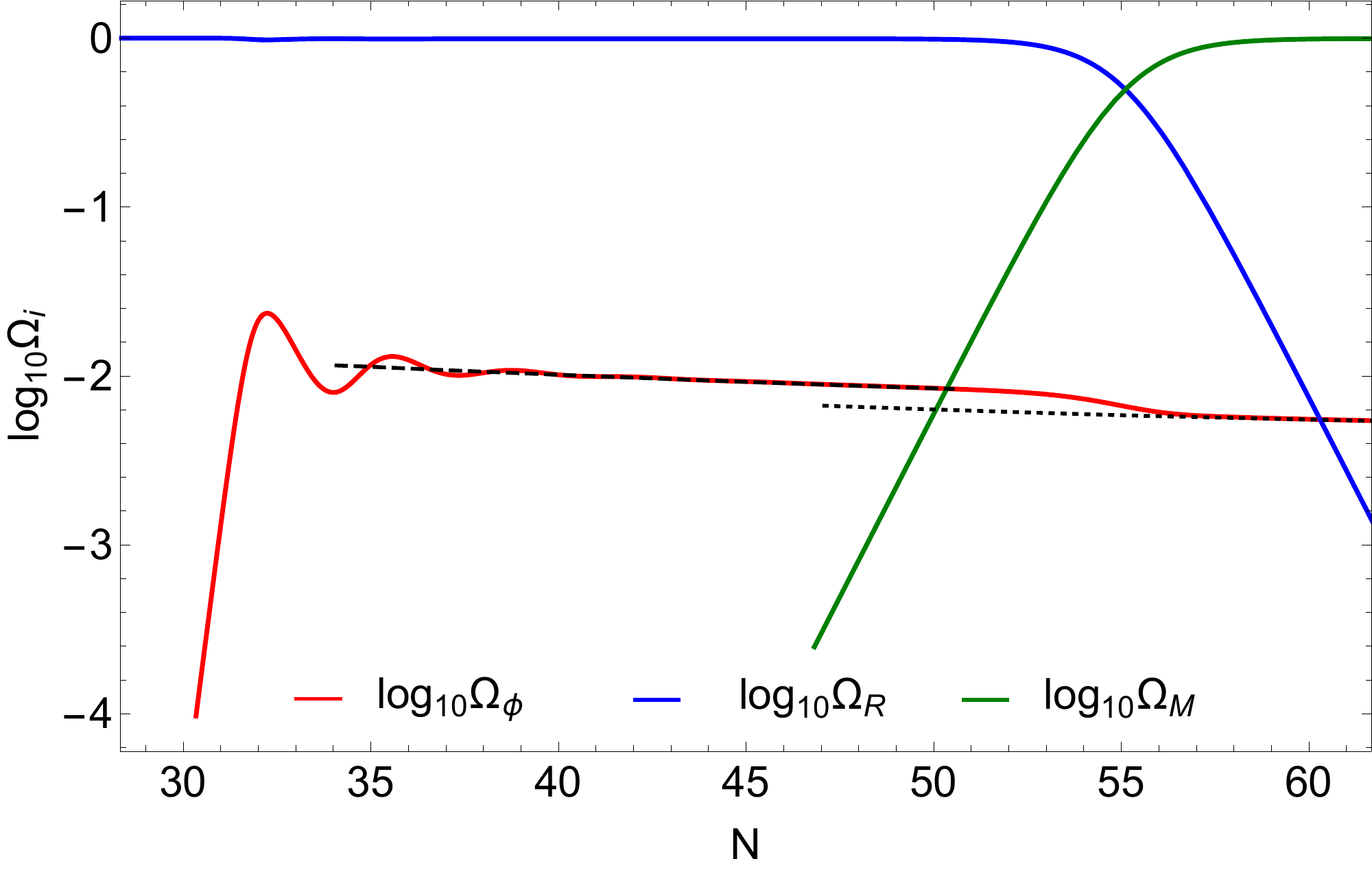}
\caption{Detailed view of the density parameters $\Omega_R$, $\Omega_M$ and $\Omega_\phi$ during 
matter and radiation domination  as a function of the number of $e$-folds using the same parameters 
in Fig.~\ref{fig:wphi}. The expression 
\eqref{Oexact} with $n=4$ and $n= 3$ is depicted with black dashed lines. Nucleosynthesis corresponds 
to $N\simeq 41$.} \label{fig:omegadetail}
\end{figure}
\begin{enumerate}[(a)]
\item  At the end of inflation the radiation component $\Omega_R$ is very small, $\Omega_R\sim\Theta$.  The cosmon evolution is dominated by its kinetic energy density. This 
domination is reflected in the dark energy equation-of-state parameter $w_\phi$, which is close to $1$ for the kinetic epoch.  
The  rapid rolling of the cosmon field down the quintessence potential translates into a substantial decrease of $V(\phi)$. Once 
$\rho_R$ becomes comparable with $\rho_\phi$, a rapid decrease of the density parameter $\Omega_\phi$ takes place.  
\item When $\rho_R$ become dominant, the Hubble parameter changes its behavior to $H=1/(2t)$. As long as the kinetic energy
 of the cosmon dominates over the potential energy,  $\ddot \phi +3 H\dot \phi \simeq 0$ , the evolution \eqref{evolkin} 
 switches to
\begin{equation}
 \phi(t)=\phi_{\rm rad}+2\dot \phi_{\rm rad} t_{\rm rad} \left(1-\sqrt{\frac{t_{\rm rad}}{t}}\right)\,,
\end{equation}
with $\phi_{\rm rad}$ and $\dot\phi_{\rm rad}$ the value of the field and its velocity at the onset of 
radiation domination $t_{\rm rad}$. 
For $t\gg t_{\rm rad}$, the field approaches the constant value $\phi_{\rm f}\simeq \phi_{\rm rad}+2\dot\phi_{\rm rad} 
t_{\rm rad}$. This \textit{freezing} of
the cosmon field translates into a substantial decrease of the cosmon kinetic energy density and an eventual 
resurgence of the potential counterpart, which is, however, subdominant with respect to the radiation 
component. During this period,  the equation-of-state parameter approaches $w_\phi\simeq-1$ and the 
dark energy fraction $\Omega_\phi$ starts to grow. 
\item Once $\rho_\phi$ approaches $\rho_R$ again, the evolution settles to a scaling or tracker solution. After some 
oscillations, the dark energy equation of state attains a nearly 
constant value $w_\phi\simeq 1/3$, which evolves toward $w_\phi\simeq 0$ after matter-radiation equality. 
In both periods, the dark energy density parameter $\Omega_\phi$  \textit{tracks} the dominant energy component. 
\end{enumerate}

\subsection{Heating efficiency}\label{sec:heatingeff}

The details of the heating process are not important for observational consequences. The only thing relevant is that
the heating terminates before the end of the kination epoch. We can then use the ratio of energy densities in radiation and the 
scalar field at any moment after the end of the heating process in order to give a generalized definition of the heating 
efficiency 
$\Theta$. During the early stages of the kinetic epoch the radiation energy density is so small that it does not influence
the cosmological evolution. For instant particle production we can compute $\rho_R/\rho_\phi$ at any given time during this 
period in terms of the scaling efficiency $\Theta$ in Eq.~\eqref{fraction},
\begin{equation}\label{XAB}
\frac{\rho_R(a)}{\rho_\phi(a)}=\Theta\left(\frac{a}{a_{\rm kin}}\right)^2\,.
\end{equation}
We can employ this expression as a practical definition of $\Theta$ for scenarios with an extended heating period. 
For any time or scale factor $a$ in the region where the heating is not efficient any longer, but radiation is still subdominant,
the heating efficiency can be defined by the radiation fraction at that moment multiplied by $(a_{\rm kin}/a)^2$. Defined 
in this way, the parameter $\Theta$ is sufficient for the description of the later evolution. The details of the heating process 
beyond the determination of $\Theta$ are not needed in practice. 

 In particular, we can use Eq.~\eqref{XAB} at the beginning of radiation domination $a=a_{\rm rad}$, 
where $\rho_\phi(a_{\rm rad})=\rho_{R}(a_{\rm rad})$ and 
 \begin{equation}\label{XAB2}
\left(\frac{a_{\rm kin}}{a_{\rm rad}}\right)^2=\Theta\,.
\end{equation}
Given a heating efficiency, the number of $e$-folds needed to explain the approximate flatness 
and homogeneity of the observable Universe is completely determined. The horizon crossing of the pivot 
scale $k_{\rm hc}$ is defined as
\begin{equation}\label{khc}
 k_{\rm hc}=a_{\rm hc} H_{\rm hc}=a_{\rm end} e^{-N(k_{\rm hc})} H_{\rm hc}\,,
\end{equation}
with $N(k_{\rm hc})$ the number of $e$-folds before the end of inflation. Assuming the kinetic regime to start immediately after the end of inflation ($a_{\rm end}\simeq a_{\rm kin}$, $\rho_{\rm end}\simeq \rho_{\rm kin}$) and taking into account the scaling of the different energy 
components, we can rewrite Eq.~\eqref{khc} as
\begin{eqnarray}\label{eqNefold}
N&=&-\ln\left(\frac{k_{\rm hc}}{a_0 H_0}\right)  \nonumber
\\ &+&\ln\left(\frac{H_{\rm hc}}{H_0}\right)+\frac{1}{4}
\ln\left(\frac{\rho_{mat}}{\rho_{\rm hc}}\right)+\ln\left(\frac{a_{mat}}{a_0}\right)  \nonumber\\
&-&\frac{1}{2}\ln\left(\frac{a_{\rm kin}}{a_{\rm rad}}\right)+\frac14 \ln\left(\frac{\rho_{\rm hc}}{\rho_{\rm kin}}\right)\,,
\end{eqnarray}
where the subindices kin, rad, mat and 0 denote respectively the onset of kinetic, radiation 
and matter dominated eras and the present cosmological epoch. Equation~\eqref{eqNefold} is universal and valid for any 
inflationary potential. The precise shape of the potential is needed to relate the energy density at the end of inflation ($\rho_{\rm end}\simeq \rho_{\rm kin}$) to the energy density at horizon crossing ($\rho_{\rm hc}$). 
Neglecting the small energy density variation  between these two epochs 
($\rho_{\rm hc}\simeq\rho_{\rm end}\simeq\rho_{\rm kin}$), we  can approximate Eq.~\eqref{eqNefold} by ($a_0=1$)
\begin{align}\label{eqNefold2}
N\simeq&-\ln\left(\frac{k_{\rm hc}}{T_0\,}\right)+\frac{1}{4}
\ln\left(\frac{\pi^2 g_{\rm mat} {\cal A}}{135}\right)  \nonumber\\
&+\frac{1}{4}\ln\, r-\frac{1}{4}\ln\Theta\,,
\end{align}
where we have made use of Eqs.~\eqref{amplitude} and \eqref{XAB2} together with the standard relations
\begin{equation}
\rho_{\rm mat}\simeq \frac{\pi^2g_{\rm mat}}{15} T_{\rm mat}^4\,,\hspace{10mm}
\frac{a_{\rm mat}}{a_0}=\frac{T_0}{T_{\rm mat}}\,.
\end{equation}
Here  $g_{\rm mat}=3.36$ stands for the the number of relativistic degrees of freedom at matter-radiation equality. 
\begin{figure}
\centering
\includegraphics[scale=0.37]{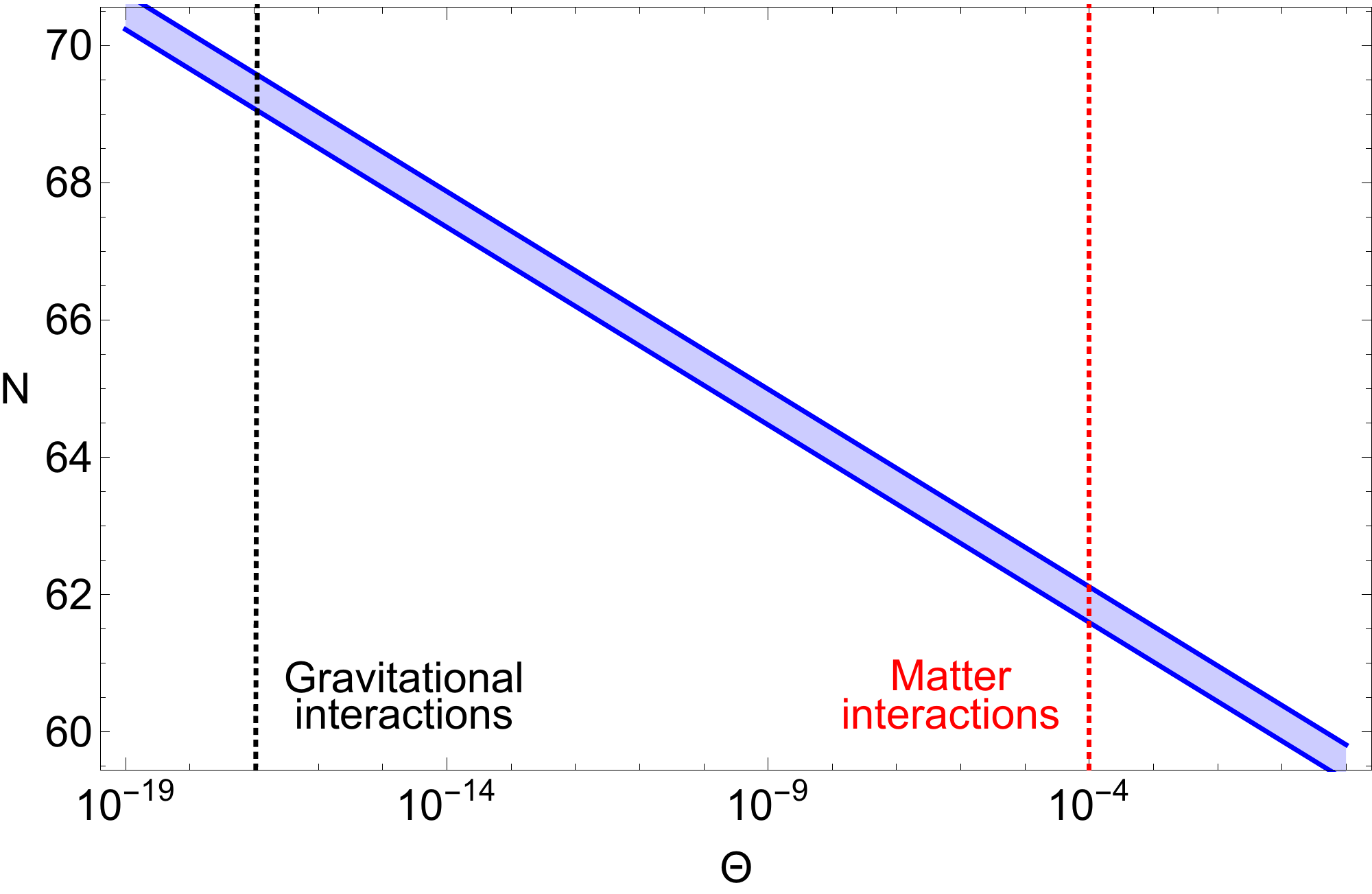}
\caption{ The number of $e$-folds \eqref{eqNefold3} as a function of the heating efficiency 
$\Theta$. The blue region stands for variations of the tensor-to-scalar ratio $r$ within the range $[0.01,0.08]$. The 
vertical lines indicate the typical values of the heating efficiency associated to gravitational and 
matter interactions. For these lines we assume a number of production channels of the order of the number 
of degrees of freedom in the Standard Model, ${\cal O}(10^2)$ (cf. Eq.~\eqref{ratioGR} and 
Table~\ref{table1}). Enhanced particle contents as those appearing in Standard Model extensions such as grand unification 
would translate into a larger heating efficiency and therefore into a smaller number of $e$-folds $N$.} \label{fig:Nefolds}
\end{figure}

Taking into account the pivot scale $k_{\rm hc}=0.002 \, {\rm Mpc}^{-1}=1.27\times 10^{-32}$ eV used in the derivation of  the \textit{combined} 
Planck/BICEP2 results \cite{Ade:2015lrj,Ade:2015xua}, Eq.~\eqref{eqNefold2} becomes\footnote{We use 
$T_0\simeq 2.73\, {\rm K}\simeq 2.35\times 10^{-4}$ eV.}
\begin{equation}\label{eqNefold3}
N\simeq62+\frac{1}{4}\ln\, \left(\frac{r}{0.05}\right)-\frac{1}{4}\ln\left(\frac{\Theta}{10^{-4}}\right)\,.
\end{equation}
The behavior of the number of $e$-folds as a function of the heating efficiency $\Theta$ is shown 
in Fig.~\ref{fig:Nefolds}.  As clearly appreciated in this figure, our highly efficient heating scenario translates 
into a rapid onset of radiation domination and into a number of $e$-folds rather close to the standard value $N \sim 60$.

\subsection{Scaling solution}

During the radiation dominated epoch the dark energy density decreases according to a scaling 
solution \cite{Wetterich:1987fm}. The behavior of the scaling solution can be  easily understood by 
considering the evolution equations for the cosmological observables during matter and radiation domination 
in terms of suitable variables.  For $\rho=\rho_R+\rho_M$, $\dot \rho + n H\dot \rho=0$ and 
constant $n$  ($n=4$ for radiation domination, $n=3$ for matter domination) one 
has \cite{Wetterich:2003qb,Scherrer:2007pu,Chiba:2012cb} 
\begin{eqnarray}
&& \hspace{-10mm}\frac{w_\phi'}{1-w_\phi}=-3(1+w_\phi)+\lambda\sqrt{3(1+w_\phi)\Omega_\phi}\label{weq}\,,  \\
&& \hspace{-10mm}\,\Omega'_\phi=-\Omega_\phi (1-\Omega_\phi)\left(3(1+w_\phi)-n \right)\,,\label{Oeq}\\ 
&& \hspace{-10mm}\lambda'=-\sqrt{3(1+w_\phi)\Omega_\phi}(\Gamma-1)\lambda^2\,,\label{Leq}
\end{eqnarray}
where the primes denote derivatives with respect to the number of $e$-folds. The slow-roll parameters 
\be\label{SRpamDE}
\lambda\equiv -M_{\rm P}\frac{V_{,\phi}}{V}\,, \hspace{10mm} \Gamma\equiv\frac{VV_{,\phi \phi}}{V_{,\phi}^2}\,,
\ee
characterize the slope and curvature of the quintessence cosmon potential. 

A simple inspection of Eqs.~\eqref{weq} and \eqref{Oeq} reveals for constant $\lambda$ a fixed point at
\begin{equation}\label{exprel}
\Omega_\phi=\frac{n}{\lambda}\,,\hspace{10mm} w_\phi=\frac{n-3}{3}\,. 
\end{equation}
This case corresponds to an exponential potential, for which a \textit{scaling} 
or \textit{tracker} solution is well known to exist 
and to be stable \cite{Wetterich:1987fm,Wetterich:1994bg,Ratra:1987rm,Copeland:1997et}.

For the potential \eqref{Wexact}, the slow-roll parameters \eqref{SRpamDE} become
\begin{equation}\label{GL}
 \lambda=2\sqrt{\frac{\kappa\, Y}{\sigma}}\,,\hspace{10mm}\Gamma=1-\frac{\sigma}{4(1+Y)}\,,
\end{equation}
with $Y$ given by Eq.~\eqref{Lamb}. Combining 
these expressions, we obtain a relation between $\Gamma$ and $\lambda$
\be
\Gamma(\lambda)=1-\frac{\kappa\sigma}{4\kappa+\sigma\lambda^2}\,,
\ee
that \textit{closes} the system of autonomous equations \eqref{weq}-\eqref{Leq}.

Since $\Gamma\neq 1$, the $\lambda$ parameter must evolve 
on time, cf. Eq.~\eqref{Leq}. Note, however, that at large field values ($\phi\gg \phi_t$), the function
$Y$ becomes approximately $Y\approx \phi^2/\phi_t^2+2\gg1$, which translates into a value of $\Gamma$ that 
tends asymptotically to one. In this limit, the relations \eqref{exprel} may still be considered as some 
\textit{fixed trajectory} with slowly varying $\lambda$. Taking into account the first equation in~\eqref{GL}, this 
asymptotic behavior can be written as
\begin{equation}\label{Oexact}
 \Omega_{\phi}= \frac{n \sigma}{4\kappa Y(\phi)}=\frac{n B(\phi(\chi))}{4}\,,   
\end{equation}
which coincides with the approximate expression at large $\chi$ found in Ref.~\cite{Wetterich:2014gaa}.

\begin{figure}
\centering
\includegraphics[scale=0.4]{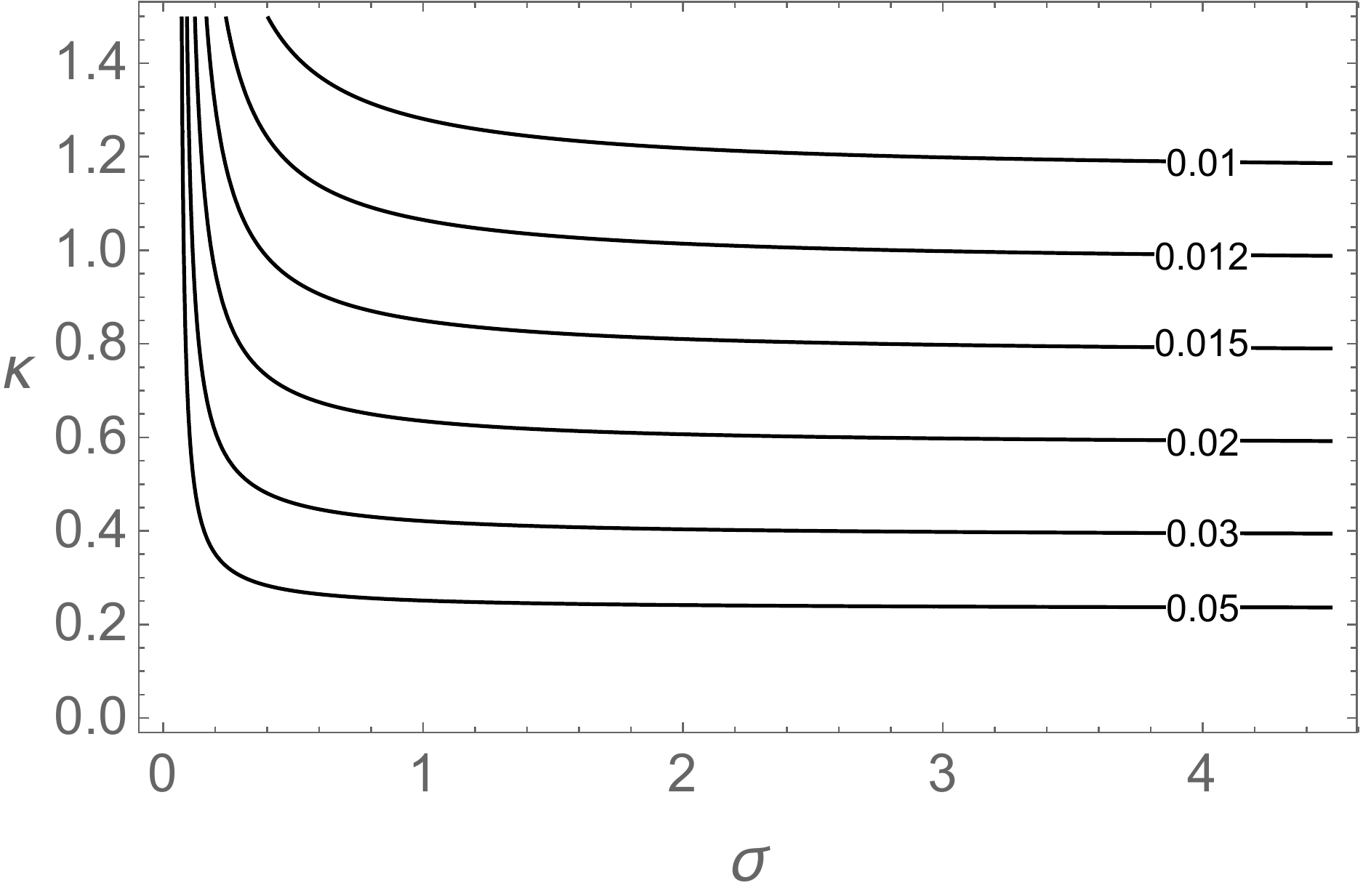}
\caption{The fraction of early dark energy \eqref{Oexact} at BBN as a function of the model 
parameters $\kappa$ and $\sigma$. For this plot we set $T_{\rm BBN}=1$ MeV and $w_\phi\simeq 1/3$.} \label{fig:kbounds}
\end{figure}

The relation \eqref{Oexact} can be used to obtain bounds on $\kappa$ from early dark energy constraints. 
Using Eq.~\eqref{Brun} we can rewrite Eq.~\eqref{Oexact} as
\begin{equation}\label{OBBN}
\Omega_{\phi}\simeq \frac{n\sigma}{4\kappa} {\cal W}^{-1}\left[\left(\frac{V_0}{V}\right)^{\sigma/2}\right]\,,
\end{equation}
with ${\cal W}$ the Lambert function and $V_0$ given by Eq.~\eqref{V0def}, with $m/\mu$ satisfying the inflationary 
constraint \eqref{movermu}. For the epoch of nucleosynthesis we may employ $M_P^4 V\approx T^4_{\rm BBN}$. The dependence
 of $\Omega_\phi^{\rm BBN}$ on $\sigma$ 
and $\kappa$ is illustrated in Fig.~\ref{fig:kbounds} for typical values $N=60$ and $T_{\rm BBN}=1$ MeV. As clearly 
seen in this figure, the early dark energy fraction strongly depends on $\kappa$, while it is rather insensitive to the 
precise value of the UV anomalous dimension $\sigma$, provided that this is not very close to zero. The current 
observations \cite{Ade:2015xua,Reichardt:2011fv,Sievers:2013ica,Pettorino:2013ia}  restrict $\Omega_\phi^{\rm BBN}$ to  be smaller than $2\%$, which can be easily satisfied for $\kappa>0.5$. 

The time when $\Omega_\phi$ reaches the scaling solution ($N\approx 32$ in Fig.~\ref{fig:omegadetail}) depends on the 
heating efficiency $\Theta$. The smaller the $\Theta$, the later the radiation domination sets in, the smaller the $V$ is when 
the evolution of the scalar field stops, and therefore the later the onset of the scaling solution is. Only after the onset 
of the scaling solution a non-negligible fraction of early dark energy is present. In particular, the BBN constraint 
applies only if the cosmon field reaches the attractor solution before BBN. This is not
guaranteed for very long kinetic regimes, even if the GW bound in Eq.~\eqref{GWbound4} is satisfied.  
For the particular parameters considered in this section,  the attractor solution is not reached 
before BBN if the heating efficiency is  below $\Theta\simeq 10^{-13}$. Further constraints on early dark energy arise from the detailed properties of the CMB spectrum, which depend on $\Omega_\phi$ at the time of CMB emission. The bounds 
\cite{Ade:2015rim} are strength similar to the BBN bounds. The presence of early dark energy during structure formation reduces the 
presently observable structure as compared to the CMB prediction in the standard cold dark matter scenario. As a rough rule, 
1\% $\Omega_\phi$ reduces $\sigma_8$ by $5\%$ \cite{Doran:2001rw}.

At the end of this section we may recall that the hot big bang picture is a property of the Einstein frame. In the scaling 
frame both the particle masses and the Planck scale increase with time. The Universe shrinks during radiation and matter
domination eras and the temperature of the Universe increases \cite{Wetterich:2014gaa,Wetterich:2013jsa,Wetterich:2013aca}. 
Only  dimensionless ratios such as temperature over particle mass or distance between galaxies over atom size show  the
same behavior in both frames.

\section{Late dark energy domination}

An exit mechanism from the scaling regime is needed in order to obtain late-time acceleration.  A rather natural 
setup arises if the neutrino-to-electron mass ratio increases with increasing $\phi$ in the present cosmological
epoch \cite{Wetterich:2007kr,Amendola:2007yx}. In our scenario this effect can be induced by a second crossover 
stage in the beyond the Standard Model sector which manifests itself through the nonrenormalizable neutrino mass operator. 
A decrease of the mass of the right-handed neutrinos or of a heavy scalar triplet (seesaw I or II mechanism) in units of $\chi$ 
results in an increase  of the mass ratio between the light left-handed neutrinos and the electron. More quantitatively, 
we may define in the scaling frame
\begin{equation}\label{Anu}
\tilde \gamma (\chi)=\frac{1}{2}\chi\frac{\partial}{\partial\chi}\ln \frac{m_{\nu}(\chi)}{\chi}\,,
\end{equation}
with $m_{\nu}(\chi)$ the average of the masses of the left-handed neutrinos. For the particular example 
\begin{equation}\label{A2nu}
m_{\nu}(\chi)=\frac{c_\nu \chi}{\ln\left(\frac{\chi_0^2}{\chi^2}\right)}\,, 
\end{equation}
one has
\begin{equation}\label{Bnu}
\tilde \gamma =\frac{1}{\ln\left(\frac{\chi_0^2}{\chi^2}\right)} \,.
\end{equation}

In the Einstein frame with fixed electron mass, only the neutrino mass depends on the cosmon field, defining 
the effective neutrino-cosmon coupling
\begin{equation}
 \beta=-M_P\frac{\partial}{\partial\phi}\ln m_{\nu}(\phi)\,.
\end{equation}
A typical behavior of this quantity, corresponding to Eq.~\eqref{A2nu}, is parametrized by 
\begin{equation}\label{Enu}
\beta=\frac{M_P}{\phi-\phi_c}\,. 
\end{equation}
Only finite values of $\beta\approx 100$ will occur, such that the singularity in Eq.~\eqref{Enu} is never reached and can be 
removed by a different behavior at small $\vert \phi-\phi_c\vert$. The cosmon-neutrino coupling modifies the Klein-Gordon 
equation for the scalar field \cite{Wetterich:1994bg,Fardon:2003eh,Brookfield:2005bz}
\begin{equation}
\ddot \phi +3 H\dot \phi=-V_{,\phi} +\frac{\beta}{M_P}\left(\rho_{\nu}-3 p_{\nu}\right)\,,
\end{equation}
and the conservation equation for the neutrino energy density
\begin{equation}
\dot\rho_{\nu}+3H(\rho_\nu+p_\nu)=-\frac{\beta}{M_P}\left(\rho_{\nu}-3p_{\nu}\right)\dot\phi\,. 
\end{equation}
These modifications are negligible as long as neutrinos are relativistic, $p_{\nu}=\rho_\nu/3$. As soon as neutrinos 
become nonrelativistic, a negative coupling $\beta$ effectively stops the evolution of $\phi$, ending the scaling solution and leading 
to a cosmology that looks rather close to a cosmological constant afterward. More precisely, the ratio of dark energy 
to neutrino energy density quickly approaches the value
\begin{equation}
 \frac{\Omega_{\phi}}{\Omega_\nu}=\tilde\gamma=-\frac{\beta}{M_P} \left(\frac{\partial \ln V}{\partial \phi}\right)^{-1}\,.
\end{equation}

Neutrinos become nonrelativistic at 
redshift $z_{\rm NR}\approx 5$ \cite{Mota:2008nj}, and the present dark energy density corresponds 
to $\Omega_\phi\,\rho_c$ at $z_{\rm NR}$. 
The resulting relation between the present dark energy density and neutrinos involves a dimensionless parameter $\gamma_\nu$ 
for the growth rate of the neutrino mass \cite{Amendola:2007yx},
\begin{equation}
\rho_\phi(t_0)^{1/4}=1.27\left(\frac{\gamma_\nu m_\nu(t_0)}{{\rm eV}}\right)^{1/4} 10^{-3}\, {\rm eV}\,,
\end{equation}
with $\gamma_\nu=\tilde \gamma(t_0)$ and $m_\nu(t_0)$ the average of the present neutrino masses. This is of 
the same order as the observed value $\rho_\phi(t_0)^{1/4}=2\cdot 10^{-3}$ eV. The equation of state is close to $-1$,
\begin{equation}
w(t_0)=-1+\frac{m_{\nu}(t_0)}{12\,{\rm eV}}\,.
\end{equation}

For $\gamma_\nu$ of order one, the scenario
is rather successful 
for the range of neutrino masses compatible with observations. Realistic setups taking into account
backreaction effects can be built on this mechanism \cite{Casas:2016duf}.

\section{Conclusions}\label{sec:conclusions}

\begin{figure}
\includegraphics[scale=1.1,center]{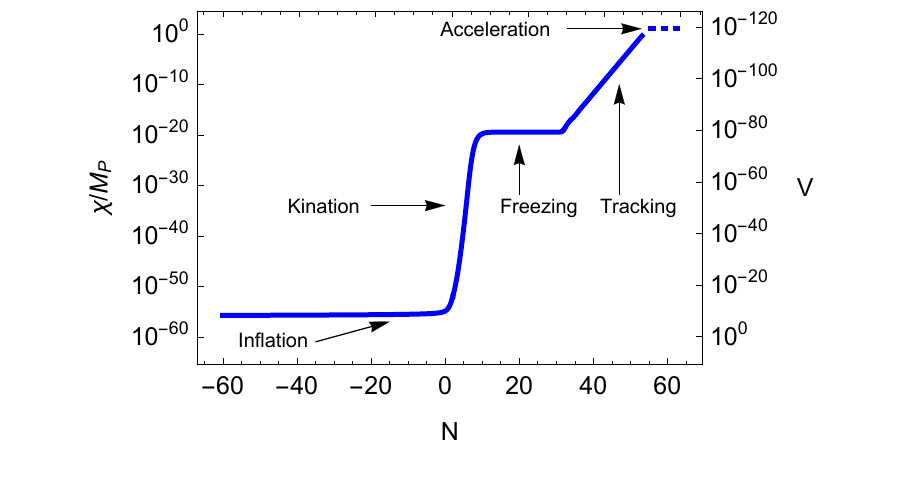}
\caption{Evolution of the cosmon field $\chi$ and the dimensionless cosmon potential \eqref{Wexact} as a function of 
the number of $e$-folds. The end of inflation corresponds to $N=0$. For 
this plot, we chose $\sigma=4$, $\kappa=1$ and assumed a heating efficiency $\Theta\simeq 10^{-4}$.} \label{fig:chievol}
\end{figure}

At the UV and IR fixed points of a variable gravity scenario scale
invariance becomes an exact symmetry of the quantum theory. Such a simple setup can remarkably 
give rise to inflation and dark energy using a \textit{single} scalar field. Approximate scale symmetry 
near the UV-fixed point manifests itself in the approximate scale invariance of the primordial fluctuation spectrum. 
Approximate scale symmetry near the IR fixed point produces an almost massless \textit{cosmon field} responsible for 
present dynamical dark energy. In the limit of exact scale symmetry this field becomes the massless Goldstone boson of 
spontaneously broken scale invariance (dilaton). A simple quadratic potential and a moderately varying kinetic term in the 
scaling frame produce a rich cosmological history, the sequence of epochs of which is summarized in Fig.~\ref{fig:chievol}. 

The graviscalar sector of our model involves only a small number of dimensionless parameters of order one: $\sigma$, 
$\kappa$, $c_t=\ln(m/\mu)$ and $\gamma_\nu$. In particular, it explains the tiny value of the present dark energy 
density (in the unit of $M_P$)  without involving any tiny or large coupling and without any tuning of couplings. If quantum gravity generates indeed an 
effective action of the type \eqref{actionJ}, this solves the cosmological constant problem. If furthermore, there exists 
a second crossover stage leading to a growing neutrino to electron mass ratio, this solves the why now problem of dark energy. The main 
 characteristics of our model are not taken completely \textit{ad hoc}. They are rather directly related to general 
 properties at the fixed points.

We derived a compact Einstein-frame 
formulation of the model in terms of Lambert functions  and proved that it can support inflation via 
a power-law inflationary potential. The spectrum of primordial fluctuations turned out to depend only on
the UV anomalous dimension $\sigma$, while the amplitude is set by the integration constant of the running kinetic term 
or the ratio $\mu/m$. The crossover to the IR fixed point manifests itself 
as a steep potential in the Einstein frame. This ends inflation and triggers the onset of a kinetic domination regime. By
considering two natural heating scenarios within the variable gravity framework, we showed that
this kination era is limited in time. The cosmologically relevant properties of the heating process can be summarized in a 
single parameter, the heating efficiency $\Theta$. Kinetic domination is naturally followed by a standard hot big 
bang era, where the (subdominant) dark energy component \textit{tracks} the (dominant) radiation/matter 
content. By comparing this tracking solution with early dark energy constraints, we derived a lower 
bound on the IR parameter $\kappa$.  The end of the scaling behavior  and the beginning  of 
the present accelerated expansion of the Universe can be induced by an additional crossover in 
sectors beyond the Standard Model. This determines the last free parameter $\gamma_\nu$ related to the present 
growth rate  of the neutrino mass. 

With all parameters determined or constrained by present observations, our model is rather predictive. We will see if its 
simplest form can survive the next round of cosmological tests.

 \section*{Acknowledgement} We acknowledge support from the ERC Advanced Grant ERC-AdG-290623 and from DFG through the 
project TRR33 ``The Dark Universe''.  

\appendix

\section{The Lambert function}\label{sec:app1}

The Lambert function ${\cal W}$ \cite{refLambert} is defined as the inverse of the
function $f(\tilde x) = \tilde x e^{\tilde x}$, i.e.
\begin{equation}
\tilde x=f^{-1}(\tilde x e^{\tilde x})={\cal W}(\tilde x e^{\tilde x})\,.
\end{equation}
Substituting $x=\tilde x e^{\tilde x}$ in this expression we obtain the defining equation for ${\cal W}(x)$
\begin{equation}\label{Lamdef}
 x={\cal W}(x)e^{{\cal W}(x)}\,.
\end{equation} 
Note that 
\begin{equation}
{\cal W}(x) +\ln {\cal W}(x) =\ln x \hspace{5mm}\textrm{for}\hspace{5mm} x>0\,.
\end{equation}
By implicit differentiation, one can show that the Lambert function satisfies 
the differential equation 
\begin{equation}
\frac{d{\cal W}}{dx}=\frac{1}{x+e^{\cal W}(x)}   \hspace{5mm}\textrm{for}\hspace{5mm} x\neq -1/e\,.
\end{equation}
Other useful relations are
\begin{equation}
 \int {\cal W} \,dx=x\,{\cal W}(x)-x+e^{{\cal W}(x)}+c\,,
\end{equation}
and ${\cal W}(0)=0$, ${\cal W}(e)=1$, or
\begin{equation}
 \int_0^e {\cal W}(x)dx=e-1\,.
\end{equation}

\section{Particle production: Analytical estimates} \label{sec:estimates}

\begin{figure}
\centering
\includegraphics[scale=0.75]{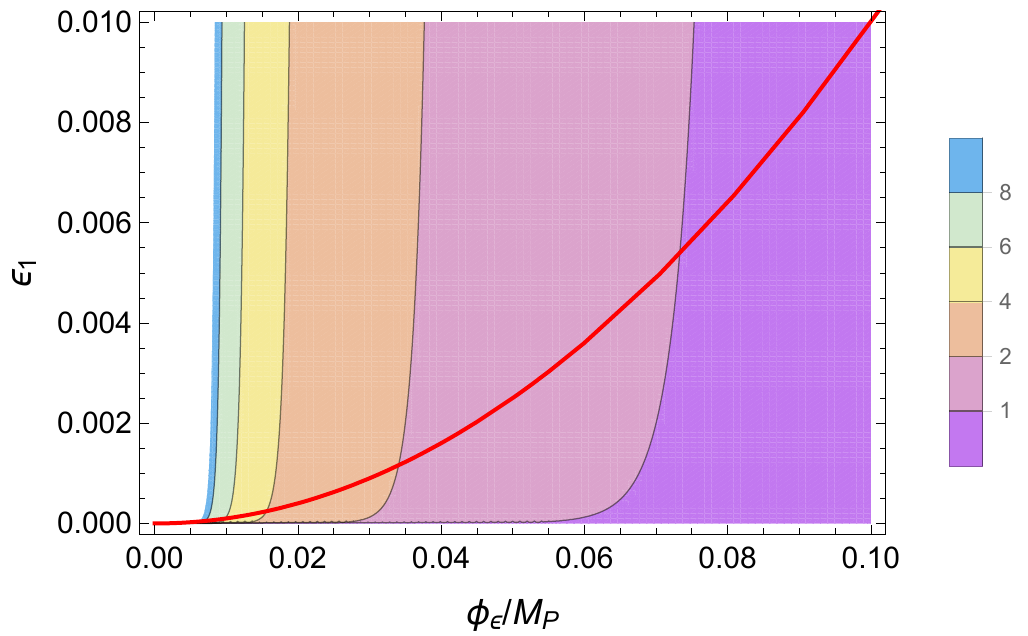}
\caption{Contour plot of the adiabaticity violation parameter $\delta_\omega$ according to Eq.~\eqref{adviol} in the 
plane of parameters $\phi_\epsilon$ and $\epsilon_1$. For the other parameters we employ $\sigma_h=2$, 
$\epsilon_\infty=10^{-10}$, a typical field velocity 
$\vert \dot \phi(\phi_{a}^{\rm max})\vert=6\times 10^{-6}M_P^2$ and $k=10^{-4}M_P$. The violation of 
adiabaticity increases toward smaller $\phi_\epsilon$. The red solid line 
corresponds to $\epsilon_1 M_P^2=\phi_\epsilon^{2}$.} \label{fig:regionAD}
\end{figure}

In this Appendix we estimate the range of parameters giving rise to significant particle production through 
the field-dependent coupling \eqref{ephi2} in Eq.~\eqref{motion}. For this purpose we evaluate 
the adiabaticity violation parameter
\begin{equation}
\delta_\omega\equiv\frac{ \vert \dot \omega_k\vert}{\omega_k^2}\,.
\end{equation}
Substancial particle production occurs if $\delta_\omega\gg 1$. For the sake of simplicity we neglect 
the expansion of the Universe. In this approximation, the adiabaticity violation parameter reads 
\begin{equation}\label{adi}
\delta_\omega=\frac{\sigma_h \vert\dot\phi\vert}{2}
 \frac{M_P^2\left(\epsilon(\phi)-\epsilon_\infty\right)}{(k^2+\epsilon(\phi) M_P^2)^{3/2}}
 \frac{(1-Y_\epsilon(\phi))}{\phi}\,.
\end{equation}
For $\epsilon_\infty\neq 0$ and/or $k\neq 0$, particle production is restricted to a compact field range. The
maximum violation of adiabaticity for a given $k$ happens at a field value $\phi_{\rm max}$ satisfying
\begin{equation}
\epsilon(\phi_{\rm max})-\epsilon_\infty=2\left(\epsilon_\infty+\frac{k^2}{M_P^2}\right) \Upsilon\,,
\end{equation}
with
\begin{equation}
\Upsilon=  1-\frac{3}{2+\sigma_h(1+Y_\epsilon(\phi_{\rm max})) }\,.
\end{equation}
This equation cannot be generically solved for $\phi_{\rm max}$. In the limit
$\sigma_h(1+Y_\epsilon(\phi_{\rm max}))\gg 1$ we can extract an approximate field value
\begin{equation} 
\frac{\phi_a^{\rm max}}{\phi_\epsilon}\approx\frac{1-Y^{\rm max}_a}{\sqrt{Y^{\rm max}_a}}\,,
\end{equation}
with $Y^{\rm max}_a$ the Lambert function 
\begin{equation}
 Y^{\rm max}_a\equiv{\cal W}\left[\left(\frac{\epsilon_1 M_P^2}{2(k^2+\epsilon_\infty M_P^2)}\right)^{2/\sigma_h}\right]\,.
\end{equation} 
At this field value, we have
\begin{equation}\label{adviol}
\delta_\omega\Big\vert_{\phi_a^{\rm max}}\approx
  \frac{\sigma_h\vert \dot \phi(\phi_{a}^{\rm max})\vert  }{3\sqrt{3}\,
  \phi_\epsilon}\sqrt{\frac{Y^{\rm max}_a}{k^2+\epsilon_\infty M_P^2}}\,\,.
\end{equation}
The values of the adiabaticity violation parameter $\delta_\omega$ for $\sigma_h=2$,  a
typical field velocity $\vert \dot \phi(\phi_{a}^{\rm max})\vert=6\times 10^{-6}M_P^2$ and $k=10^{-4}M_P$ 
are illustrated in Fig.~\ref{fig:regionAD}.. The  red solid line corresponds 
to $\epsilon_1 M_P^2=\phi_\epsilon^{2}$. As clearly 
appreciated in this figure, a significant production of highly energetic particles requires small  
values of $\phi_\epsilon/M_P$. 

\section{Cosmon production}\label{sec:app3}

As any other particle coupled to the background field $\phi$, cosmon excitations can be created via violations of the 
adiabaticity condition in the heating stage after inflation. If significantly produced, the small mass of the 
cosmon makes it a potential candidate for contributing to the effective number of relativistic degrees of freedom at BBN 
and later.  

 In Fourier space, the cosmon perturbations  $\delta\phi$ satisfy a mode equation
\begin{equation}\label{cosmonprod}
\ddot\delta\phi_k+\left(\frac{k^2}{a^2}+ M_c^2(\phi)\right)\delta\phi_k=0\,.
\end{equation}
with $M_c^2(\phi)=M_P^4V_{,\phi\phi}$ an effective mass term constructed out of the exact 
cosmon potential \eqref{Wexact}. During the inflationary stage the main contribution to this mass term 
is given by Eq.~\eqref{MCapprox}. Due to the crossover it changes more rapidly at the end of inflation and 
in the early kinetic domination period. From 
Eq.~\eqref{cosmonprod}, the energy density transferred into cosmon excitations 
at the onset of the kinetic regime can be computed
by the techniques presented in Section \ref{subsec:gravrh}. We obtain 
\begin{equation}\label{cosmonP}
\frac{\rho_{\delta\phi}^{\rm kin}}{\rho^{\rm kin}_\phi}\simeq 7\,\times 10^{-17} \,.
\end{equation}
Taking into account the results in Table \ref{table1} and associating them to the production of Standard Model
degrees of freedom ($\rho_{\rm h}^{\rm kin}\simeq \rho^{\rm kin}_{\rm SM}$), Eq.~\eqref{cosmonP} translates into a ratio
\begin{equation}\label{Crange}
3\times \,10^{-8}\geq \frac{\rho_{\delta\phi}^{\rm kin}}{\rho^{\rm kin}_\textrm{SM}}\geq 5.3\times \, 10^{-11}\,,
\end{equation}
 for values of $\phi_\epsilon$ in the range $10^{-4}\leq \phi_\epsilon\leq 10^{-1}$. 
 
Additional relativistic degrees of freedom on top of the Standard Model ones are typically parametrized in terms of an
effective number of neutrinos at BBN
 \begin{equation}\label{Neff}
 \Delta N_\textrm{eff}\equiv\frac{\rho^{\rm BBN}_\phi}{\rho^{\rm BBN}_\nu}\,,
  \end{equation}
with  $\rho_{\nu}=\frac{\pi^2}{30}g_{\nu}T_{\textrm{f}}^4$ the energy density associated to a single neutrino species.  
 Assuming complete thermalization at the onset of kinetic domination and taking into account the scaling 
 of the different components, this quantity can be easily related to  $\rho_{\delta\phi}^{\rm kin}/\rho^{\rm kin}_\textrm{SM}$ up to an order one numerical factor associated to the change of relativistic degrees of freedom from the onset of kinetic domination to the BBN era (for details see Ref.~\cite{GarciaBellido:2012zu})
   \begin{equation}\label{Neff2}
 \Delta N_\textrm{eff}\propto \frac{\rho_{\delta\phi}^{\rm kin}}{\rho^{\rm kin}_\textrm{SM}}\,.
  \end{equation}
We conclude therefore that the contribution of cosmon excitations \eqref{Crange} to the effective 
number of neutrinos at BBN is tiny, well within the 
cosmological bound $\Delta N_{\rm eff}\lesssim  0.15\pm 0.23$ provided by the Planck 
Collaboration \cite{Ade:2015xua}. The cosmon  remains an elusive particle that cannot be detected by 
any astrophysical or particle physics experiment. Only its field value at large scales, from the horizon perhaps down 
to cluster scales, is accessible to observation. 

\bibliographystyle{plain}

\end{document}